\documentclass[%
 preprint,
 amsmath,amssymb,
 aps,
]{revtex4-2}

\usepackage{graphicx}
\usepackage{dcolumn}
\usepackage{bm}

\begin{document}

\preprint{}

\title{Confinement-controlled chase–escape dynamics}

\author{R. G. Rossatto}
\affiliation{%
Departamento de Física, Instituto de Física e Matemática,  Universidade Federal de Pelotas, Pelotas, Brasil}%

\author{H. A. Alvarez}%
\affiliation{Instituto de Física de Líquidos y Sistemas Biológicos, CONICET-UNLP, La Plara, Argentina }%
\affiliation{Instituto de Ciencias de la Salud, Universidad Nacional Arturo Jauretche,Florencio Varela, Argetina }%
\affiliation{Departamento de Ciencias Biológicas, Facultad de Ciencias Exactas, Universidad Nacional de La Plata, La Plata, Argentina}%

\author{C. M. Carlevaro}
\affiliation{Instituto de Física de Líquidos y Sistemas Biológicos, CONICET-UNLP, La Plara, Argentina }%
\affiliation{
Universidad Tecnológica Nacional, Facultad Regional La Plata, Centro de Investigación en Mecánica Experimental y Computacional, Berisso, Argentina
}%
\author{José Rafael Bordin}
\affiliation{Departamento de Física, Instituto de Física e Matemática,  Universidade Federal de Pelotas, Pelotas, Brasil}%
\affiliation{Fachbereich Physik, Universitat Konstanz, Germany}

\date{\today}

\begin{abstract}
We investigate a minimal chase--and--escape model on a two-dimensional square lattice with randomly distributed static obstacles, focusing on how geometric disorder controls collective pursuit dynamics. Chasers and escapers move according to short-range sensing rules, while the density of obstacles tunes the connectivity of the accessible space. Using a combination of geometric analysis, dynamical observables, survival statistics, and transport characterization, we establish a direct link between lattice connectivity and pursuit efficiency. A Breadth-First Search analysis reveals that obstacle-induced fragmentation leads to a progressive loss of accessibility before the percolation threshold, defining the effective initial conditions for the dynamics. The trapping time and capture cost exhibit a non-monotonic dependence on obstacle density, reflecting a competition between path elongation in connected environments and geometric confinement near the percolation threshold. Survival analysis shows that the decay of the escaper population follows a Weibull form, with characteristic time and shape parameters displaying clear crossovers as a function of obstacle density, signaling the coexistence of cooperative capture and confinement-dominated trapping. Transport properties, quantified through the mean-squared displacement exponent, further support this picture, revealing sub-diffusive dynamics and a convergence toward a geometry-controlled regime near percolation. Overall, our results demonstrate that chase--and--escape dynamics in disordered environments are governed by a geometry-driven crossover, where percolation and connectivity act as unifying control parameters for spatial, temporal, and collective behavior.

\end{abstract}

\maketitle


\section{Introduction}
\label{introd}

Chase--and--escape dynamics have attracted sustained interest in statistical physics, soft and active matter, and complex systems as minimal models linking local interaction rules to emergent collective behavior~\cite{kamimura2010NJP,Angelani2012,Vicsek2012,Janosov2017,Ohira2024}. In active and driven systems, geometry and crowding play a central role in shaping transport, encounter rates, and coordination, making pursuit--evasion processes a natural testbed for non-equilibrium organization~\cite{RevModPhys.88.045006,Bowick22,Epstein2019,teVrugt2025}. These models also provide minimal realizations of non-equilibrium transitions between active and absorbing states, with close connections to epidemic spreading, diffusion--limited reactions, and predator--prey dynamics~\cite{RevModPhys.88.045006,Bowick22,evans1993RMP}.

While many theoretical formulations assume open or weakly structured environments, realistic scenarios are strongly constrained by obstacles and heterogeneous connectivity. Such constraints are ubiquitous in biological systems, where immune cells navigate complex tissues~\cite{Fowell2021,Schienstock2021,MoreiraSoares2020,Melo2023}, in fragmented ecological landscapes~\cite{Gorini11,Cozzi2013}, and in technological applications such as swarm robotics and autonomous navigation in cluttered environments~\cite{Oyler2016,Liang2023,Katona2024,Zhou2021,Yaacoub2021,Alqudsi2025}. Related crowd and evacuation dynamics further illustrate how boundaries and bottlenecks control collective motion and escape efficiency~\cite{helbing2005TS,zhang2021PRA,wang2025JCSS}.

From a statistical--physics perspective, disorder and confinement naturally induce anomalous transport, subdiffusion, and trapping, phenomena widely observed in active and heterogeneous media~\cite{RevModPhys.88.045006,chepizhko13,Bellomo2020}. Geometric constraints generate long-time correlations and broad distributions of trapping times, leading to non-Gaussian dynamics and heavy-tailed survival statistics. Lattice-based models provide a controlled framework to disentangle geometric and dynamical effects, establishing a direct connection between spatial accessibility, percolation, and pursuit efficiency~\cite{Kumar2021,cornette2003Springer}. In particular, the introduction of static obstacles in two-dimensional chase--escape models gives rise to nontrivial survival curves and non-monotonic efficiency trade--offs~\cite{vscepanovic2019PhysicaA}, while related graph-based approaches capture finite-connectivity effects and phase transitions in pursuit processes~\cite{Fortunato2010,NewmanGirvan2004,bernstein2022ECP,hinsen2019arvix,cali2024JAP,Beckman2021}.

Motivated by these considerations, we investigate a minimal chase--and--escape model on a square lattice with randomly distributed static obstacles, where chasers and escapers move according to short-range sensing rules and stochastic updates~\cite{wang2017PhysicaA,su2023NJP}. By systematically varying obstacle density and population ratios, we analyze how geometric disorder controls (i) the trapping time as a measure of global efficiency, (ii) a cost function associated with collective effort, (iii) survival statistics through the survival function $S(t)$ and the cumulative hazard $H(t)$, and (iv) transport anomalies quantified by the mean-square displacement (MSD) exponent $\alpha$.

Our results reveal a geometry-driven crossover in the pursuit dynamics. Intermediate obstacle densities maximize trapping time and cost, reflecting hindered exploration in still-connected domains, whereas beyond the connectivity breakdown geometric fragmentation dominates and reduces both efficiency and variability. The decay of the escaper population follows a Weibull distribution, indicating two distinct capture regimes: an early-time compressed exponential associated with cooperative pursuit and a long-time stretched exponential corresponding to confinement and geometric trapping. This crossover is a hallmark of subdiffusive, trap-dominated dynamics in heterogeneous media~\cite{RevModPhys.88.045006,chepizhko13}. By integrating percolation analysis, survival statistics, and transport characterization, this work establishes a unified statistical--physics description of cooperative pursuit in disordered environments.

The remainder of this article is organized as follows. In Section~2, we describe the model and simulation protocol. Section~3 presents and discusses the main results. Finally, Section~4 summarizes the conclusions and outlines perspectives for future research.

\section{Model and Computational Methods}

We consider a chase--and--escape system defined on a two-dimensional square lattice of linear size $L = 128$ (results for other system sizes are reported in the Supplementary Material). The system consists of two types of agents: $N^{\rm C}$ chasers (hunters) and $N^{\rm E}$ escapers (prey), as well as static obstacles distributed over the lattice. Periodic boundary conditions are employed in all simulations.

Obstacles are placed first using a random sequential adsorption (RSA) protocol~\cite{evans1993RMP}. The obstacle density is defined as $\phi = N_{\rm obs}/L^{2}$, where $N_{\rm obs}$ is the number of obstacles, and takes values in the range $\phi \in \{0.00, 0.10, 0.20,\\ 0.30, 0.40, 0.50, 0.59, 0.60, 0.61, 0.70, 0.80\}$. The system undergoes a geometric percolation transition at a critical obstacle density $\phi_c$. For site percolation on a square lattice, the critical threshold in the thermodynamic limit is $\phi_c \simeq 0.593$~\cite{cornette2003Springer}. In the finite system considered here, we define an effective percolation threshold $\phi_c = 0.60$.

The initial number of escapers is defined as a fixed fraction of the free lattice sites, as well the number of chasers is a function of the number of escapers, 
\begin{equation}
N^{\rm E}_0 = \frac{L^2 (1-\phi)}{4},
\end{equation}
\begin{equation}
N^{\rm C} = f N^{\rm E}_0,
\end{equation}
with $f \in \{0.5, 0.8, 1.0\}$. Birth and death processes are not included.

Agent dynamics is modeled using an agent-based approach with discrete-time updates. Both chasers and escapers interact through a finite sensing or search radius ($SR$), inspired by Ref.~\cite{vscepanovic2019PhysicaA} and simplified relative to Refs.~\cite{su2023NJP,wang2017PhysicaA}. The sensing radius is defined using the Manhattan metric: the distance between two lattice sites $S_1=(x_1,y_1)$ and $S_2=(x_2,y_2)$ is given by
\begin{equation}
d_{\rm M} = |x_1-x_2| + |y_1-y_2|.
\end{equation}
An agent detects another agent if $d_{\rm M} \leq SR$, with $SR \in \{1,2\}$.

\begin{figure}[h!]
    \centering
    \includegraphics[width=0.8\textwidth]{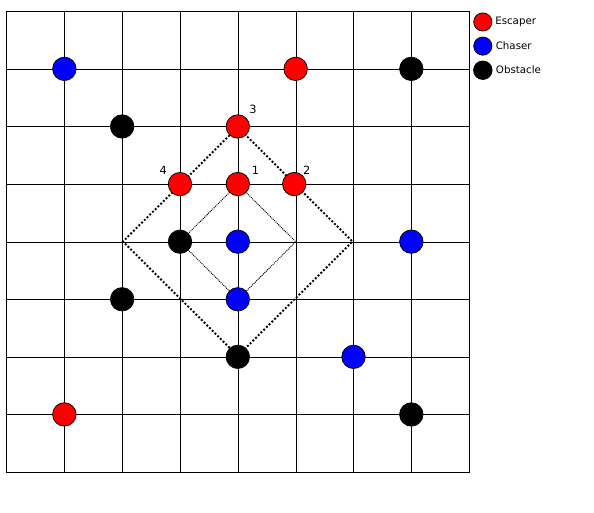}
\caption{Schematic illustration of the sensing radius (SR) defined using the Manhattan metric for a chaser located at the center of the lattice. Sites within SR = 1 and SR = 2 determine the regions where directed motion rules apply. This schematic illustrates the local decision rules used in the agent-based dynamics and is not drawn to scale.}
    \label{Ma.Dis}
\end{figure}

At each simulation step, a lattice site is selected at random. If the site is empty or occupied by an obstacle, no update occurs. If the site contains an agent, it attempts to move according to the following rules:
(i) chasers move toward the nearest escaper detected within $SR=1$, choosing randomly if multiple targets are available;
(ii) if no escaper is detected within $SR=1$ but at least one is detected within $SR=2$, the chaser moves in the direction of the highest local escaper density;
(iii) if no escaper is detected within $SR=2$, the chaser performs a random walk.
Escapers follow analogous rules, moving away from nearby chasers within $SR=1$ or from regions of high chaser density within $SR=2$, and performing a random walk otherwise.
A schematic illustration of the sensing radius is shown in Fig.~\ref{Ma.Dis}.

A central quantity characterizing the pursuit efficiency is the trapping time $TT$, defined as the number of simulation steps required to capture all accessible escapers. Following Ref.~\cite{kamimura2010NJP}, we define the capture cost as
\begin{equation}
c = \frac{N^{\rm C} TT}{N^{\rm E}_0} = f \, TT,
\label{cost_function_o}
\end{equation}
which measures the collective effort required per initial escaper. This cost represents an operational measure of collective effort rather than an energetic quantity.

Survival statistics are analyzed using nonparametric survival analysis techniques. Since simulations terminate once all accessible escapers are captured and different realizations may have different durations, we employ the Kaplan--Meier estimator~\cite{kaplan1958TeF}. The survival function is defined as
\begin{equation}
S(t) = \prod_{i: t_i \le t} \left(1 - \frac{d_i}{n_i}\right),
\label{survival_function}
\end{equation}
where $d_i$ is the number of escapers captured at time $t_i$ and $n_i$ is the number of escapers at risk immediately before $t_i$. The survival function can equivalently be expressed as
\begin{equation}
S(t) = \exp[-H(t)],
\end{equation}
where $H(t)$ is the cumulative hazard function, estimated using the Nelson--Aalen estimator~\cite{colosimo2002TF},
\begin{equation}
H(t) = \sum_{i: t_i \le t} \frac{d_i}{n_i}.
\label{hazard}
\end{equation}

Transport properties are characterized through the mean-squared displacement (MSD) of the chasers, computed using the \texttt{Trajpy} package~\cite{moreira2024trajpy}. The MSD is defined as
\begin{equation}
\langle \mathbf{r}^2(t) \rangle \propto t^{\alpha},
\label{msd}
\end{equation}
where the exponent $\alpha$ characterizes the dynamical regime of the system. The MSD of chasers is used as a proxy for the global transport properties of the system, since chasers actively explore the accessible space throughout the pursuit process.

\section{Results}

\subsection{Geometric accessibility and BFS analysis}

Before performing the dynamical chase--escape simulations, we carried out a purely geometric analysis of the lattice connectivity using the Breadth-First Search (BFS) algorithm~\cite{cormen2022MIT}. The purpose of this preliminary step is to quantify how the obstacle density $\phi$ controls the topological accessibility of escapers, independently of the specific dynamical rules. Since pursuit trajectories are constrained by the connectivity of the free lattice sites, this analysis provides a geometric baseline for interpreting all subsequent dynamical observables.

\begin{figure}[h!]
    \centering
    \includegraphics[width=0.8\textwidth]{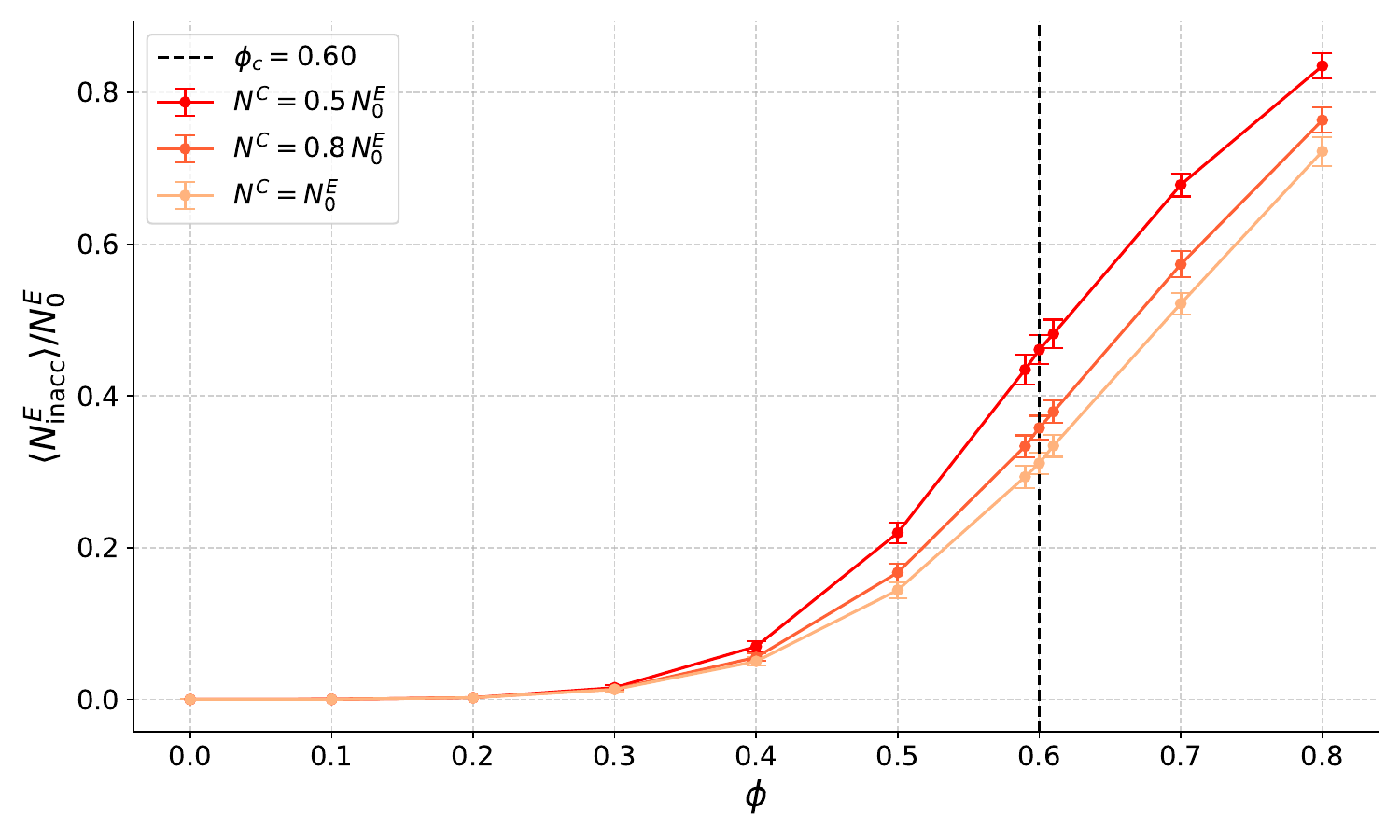}
    \caption{Fraction of topologically inaccessible escapers as a function of obstacle density $\phi$, obtained from the BFS analysis. The rapid increase observed for $\phi \gtrsim 0.50$ indicates the onset of connectivity loss prior to the percolation threshold $\phi_c$, highlighting finite-size fragmentation of the accessible space.}
    \label{I.P}
\end{figure}

Using BFS, we compute the fraction of escapers that are topologically inaccessible, \emph{i.e.}, escapers that cannot be reached by any chaser due to obstacle-induced fragmentation of the lattice. This quantity is evaluated as a function of $\phi$ for different initial chaser populations $N^{\rm C}$ and is shown in Fig.~\ref{I.P}. The dashed line indicates the percolation threshold of the obstacle network, $\phi_c$.

As $\phi$ increases, the fraction of inaccessible escapers grows monotonically, reflecting the progressive loss of global connectivity. Notably, a pronounced increase in the slope of the curves appears already for $\phi \gtrsim 0.50$, well below the theoretical site-percolation threshold in the thermodynamic limit. This behavior is a direct consequence of finite-size effects in the lattice ($L=128$), where local clusters of obstacles can fragment the accessible space before a system-spanning cluster is formed. As a result, escapers become confined within isolated pockets that are disconnected from the regions explored by chasers. These pockets define the effective initial conditions for the dynamical trapping process analyzed in the following sections. Similar finite-size precursors of percolation have been reported in lattice percolation studies~\cite{cornette2003Springer} and in chase--escape models displaying connectivity-driven transitions~\cite{Kumar2021,vscepanovic2019PhysicaA}.

\begin{figure}[h!]
    \centering
    \includegraphics[width=0.8\textwidth]{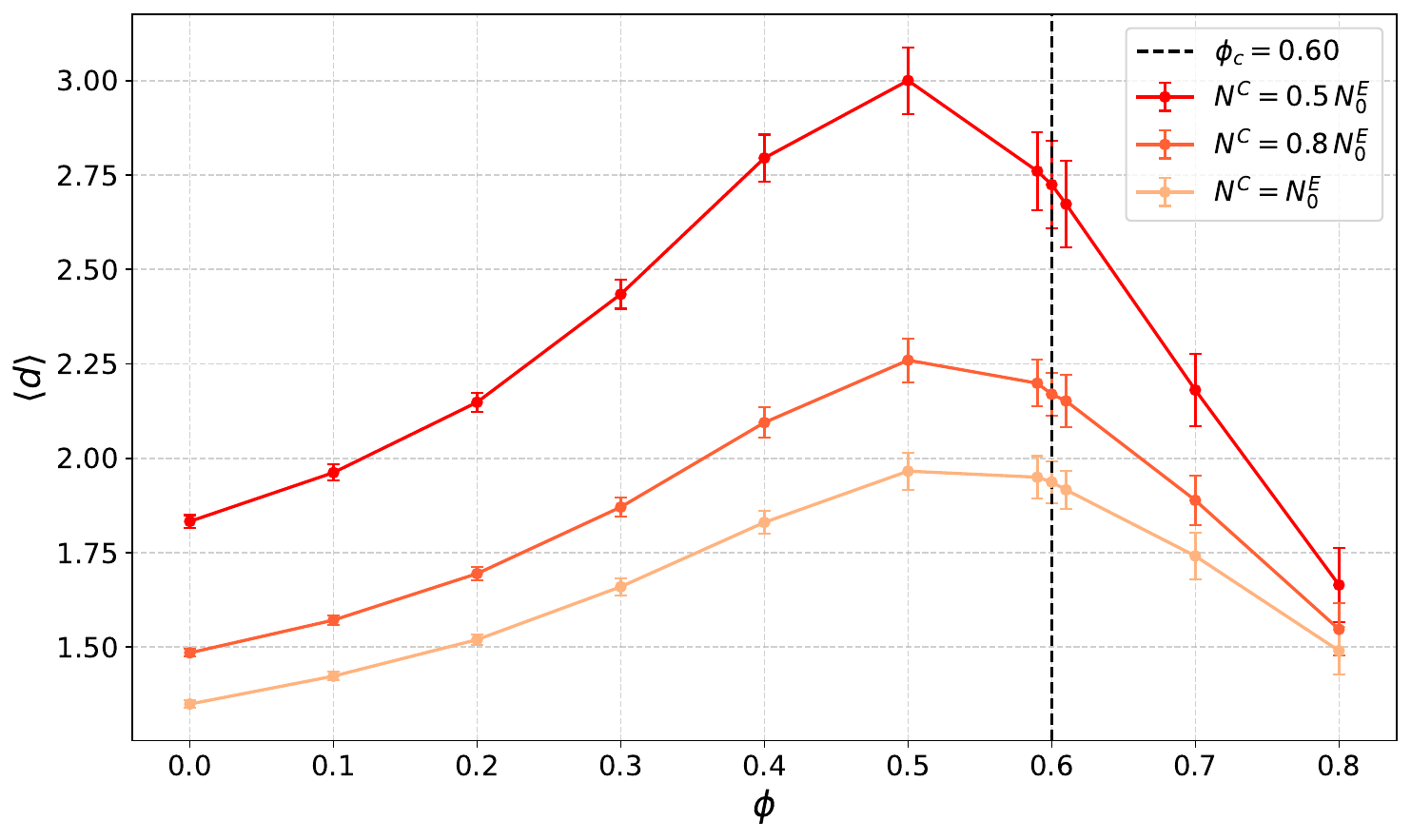}
    \caption{Mean shortest path length $\langle d \rangle$ between chasers and escapers as a function of obstacle density $\phi$, computed using BFS and considering only topologically accessible pairs. The nonmonotonic behavior reflects a crossover from path elongation in connected lattices to co-confinement within isolated clusters at high $\phi$.}
    \label{d_x_phi}
\end{figure}

Beyond accessibility, the BFS analysis also allows us to quantify characteristic geometric length scales of the system. Specifically, we compute the mean shortest path length $\langle d \rangle$ between chasers and escapers, considering only pairs for which a connecting path exists. This measure captures the effective geometric separation relevant for pursuit dynamics and is shown in Fig.~\ref{d_x_phi} as a function of $\phi$.

At low obstacle densities, $\langle d \rangle$ increases with $\phi$, reaching a maximum around $\phi \approx 0.50$. In this regime, the introduction of obstacles elongates paths within a still-connected lattice, increasing the typical distance that chasers must traverse to reach escapers. For larger $\phi$, however, $\langle d \rangle$ decreases sharply. This inversion signals a qualitative change in the geometry of the accessible space: chasers and escapers become increasingly confined within the same small clusters of free sites, where local spatial correlations dominate over long-range connectivity. For $\phi > \phi_c$, the curves corresponding to different values of $N^{\rm C}$ collapse, indicating that the geometry of the lattice, rather than the relative agent densities, becomes the dominant factor controlling the system.

This geometric crossover anticipates the dynamical regimes discussed later. In particular, the reduction of $\langle d \rangle$ at high $\phi$ reflects strong confinement and geometric trapping, which ultimately governs transport properties and capture efficiency. Such geometry-dominated behavior is consistent with observations in active particle systems moving in heterogeneous and crowded environments~\cite{RevModPhys.88.045006,chepizhko13}.

\subsection{Trapping time and geometric slowing down}

\begin{figure}[h!]
    \centering
    \includegraphics[width=0.8\textwidth]{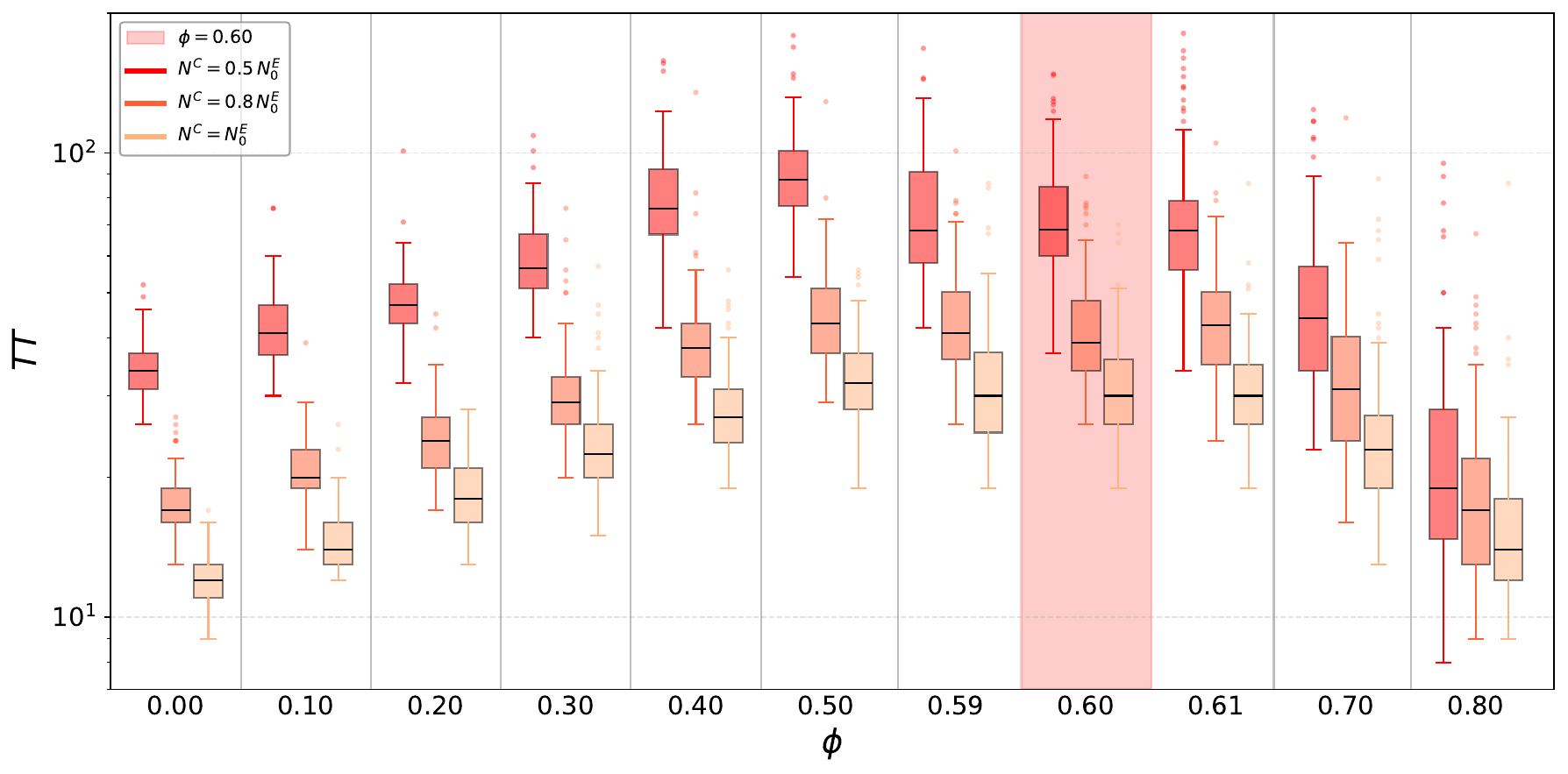}
    \caption{Trapping time $TT$ as a function of obstacle density $\phi$ for different initial chaser populations. The nonmonotonic behavior reflects a competition between path elongation in connected lattices and geometric fragmentation near the percolation regime. The shaded region marks the effective percolation threshold $\phi_c$.}
    \label{tt_x_phi}
\end{figure}

The trapping time ($TT$) provides a global measure of the dynamical efficiency of the pursuit process, defined as the total number of simulation steps required for chasers to capture all \emph{topologically accessible} escapers. As such, $TT$ directly reflects how the geometric constraints identified by the BFS analysis modulate the collective dynamics. In particular, changes in lattice connectivity and characteristic path lengths, discussed in the previous subsection, are expected to leave clear signatures on the temporal scale of the capture process.

Figure~\ref{tt_x_phi} shows the trapping time as a function of the obstacle density $\phi$ for different initial chaser populations. Data are presented as box plots, with the percolation threshold $\phi_c$ indicated by a shaded region. For all values of $N^{\rm C}$, the median trapping time increases with $\phi$ at low and intermediate obstacle densities, reaching a pronounced maximum around $\phi \approx 0.50$. This nonmonotonic behavior arises from the competition between two opposing mechanisms. On the one hand, increasing $\phi$ hinders agent mobility by elongating accessible paths, as evidenced by the growth of the mean chaser--escaper distance shown in Fig.~\ref{d_x_phi}, thereby prolonging the pursuit. On the other hand, as $\phi$ approaches the percolation regime, the accessible space fragments into disconnected domains, reducing the effective search space and accelerating the completion of the capture process.

The sharp increase of $TT$ near $\phi \approx 0.50$ is reminiscent of a geometry-induced slowing down reminiscent of critical behavior, where exploration times grow rapidly as the lattice connectivity approaches breakdown. Similar percolation-driven crossovers have been reported in active and lattice-based systems subject to geometric disorder~\cite{RevModPhys.88.045006,chepizhko13,Kumar2021,cornette2003Springer,vscepanovic2019PhysicaA}. The location of the maximum slightly below the theoretical percolation threshold ($\phi_c \approx 0.60$) reflects finite-size effects: in a finite lattice, local clusters of obstacles can effectively percolate and fragment the accessible domain before the onset of a system-spanning cluster.

Beyond the behavior of the median, the box plots reveal how obstacle-induced fragmentation reshapes the variability of the trapping dynamics. For $\phi \lesssim 0.50$, the distributions corresponding to different chaser densities are clearly separated, indicating that agent density plays a significant role in determining pursuit efficiency in well-connected environments. Near $\phi = 0.59$ and $0.60$, partial overlap emerges between the intermediate- and high-density cases, signaling increased variability as the system approaches the percolation regime. For $\phi \gtrsim 0.70$, the distributions increasingly overlap for all values of $N^{\rm C}$, and at $\phi = 0.80$ they become nearly indistinguishable. This collapse demonstrates that, in highly fragmented environments, the trapping time becomes largely insensitive to the number of chasers and is instead controlled by the geometry of the accessible clusters. In this regime, collective pursuit is quicker but also more predictable, consistent with the emergence of geometry-dominated dynamics in confined active systems~\cite{RevModPhys.88.045006,Bowick22}.

\subsection{Cost of capture and cooperative interference}

\begin{figure}[h!]
    \centering
    \includegraphics[width=0.8\textwidth]{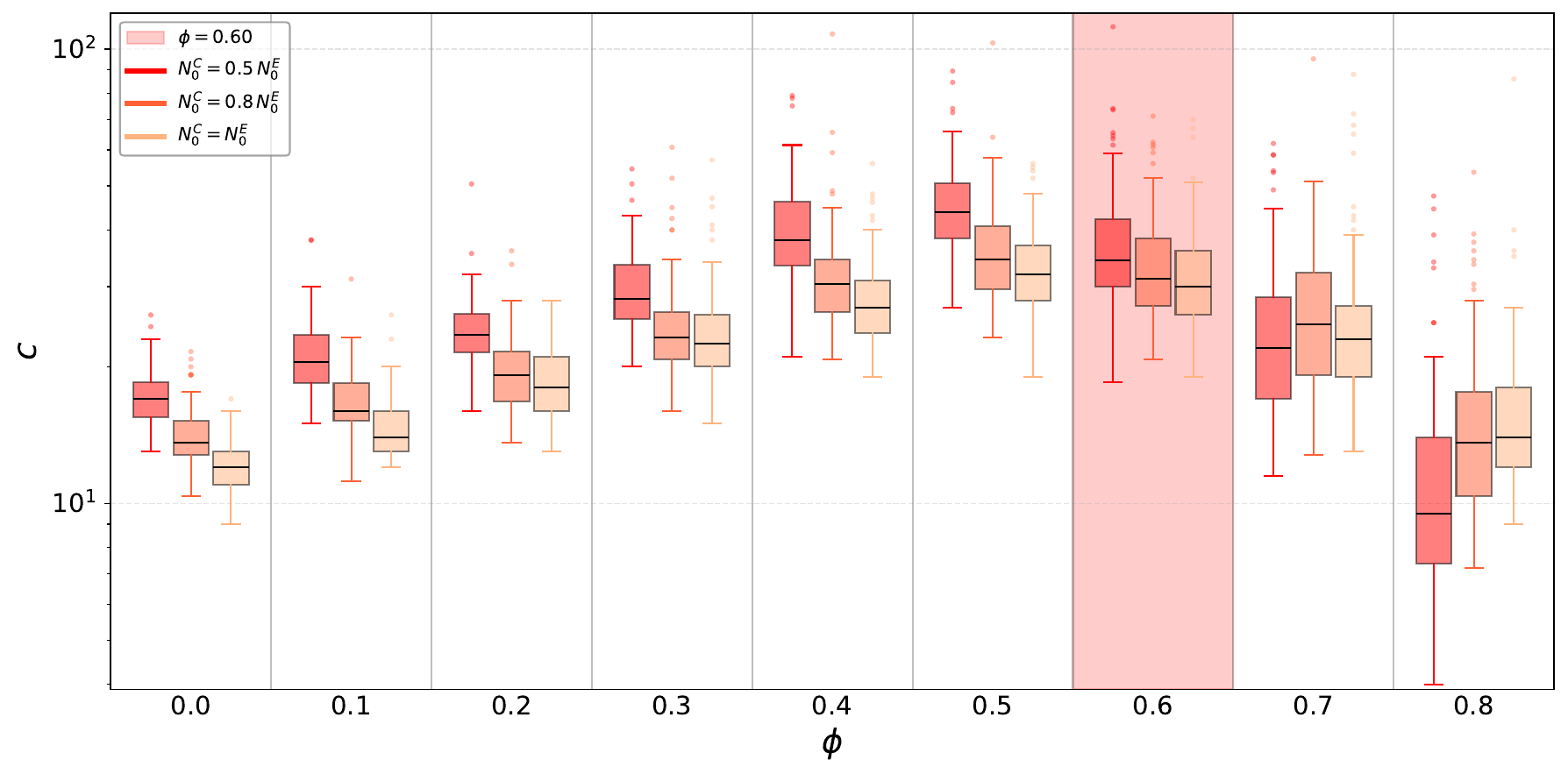}
    \caption{Cost of capture as a function of obstacle density $\phi$ for different initial chaser populations. At low $\phi$, the cost depends strongly on the number of chasers, whereas for $\phi \gtrsim \phi_c$ the distributions converge, indicating that geometric confinement dominates the pursuit efficiency.}
    \label{cost_x_phi}
\end{figure}

We now analyze the cost associated with capturing all topologically accessible escapers, shown in Fig.~\ref{cost_x_phi}. The cost, defined in Eq.~\eqref{cost_function_o} as the trapping time multiplied by the number of chasers and normalized by the initial number of escapers, provides a measure of the collective operational effort required for the pursuit process. By construction, this quantity couples the geometric constraints of the environment, encoded in the trapping time, with the population of active agents.

Figure~\ref{cost_x_phi} displays the cost as a function of obstacle density $\phi$ for different initial chaser populations. At low obstacle densities, the box plots corresponding to $N^{\rm C}=0.5N^{\rm E}_0$, $0.8N^{\rm E}_0$, and $N^{\rm E}_0$ are clearly separated, indicating that increasing the number of chasers leads to a higher operational cost in relatively open environments. As $\phi$ increases, however, these distributions progressively overlap. For $\phi \gtrsim 0.10$, partial overlap already emerges between the intermediate- and high-density cases, and near $\phi \approx 0.59$ the three distributions become largely indistinguishable. This convergence indicates that beyond a threshold obstacle density, the system cost becomes weakly dependent on the number of chasers and is instead dominated by geometric constraints imposed by spatial fragmentation.

The nonmonotonic behavior of the cost reflects the same competition observed for the trapping time: increasing obstacle density initially hinders agent motion, raising the cost, while further fragmentation near the percolation regime reduces the effective search space and limits the benefit of adding more chasers. Similar efficiency trade-offs induced by environmental constraints have been reported in multi-agent pursuit and swarm robotics models, where obstacle-induced fragmentation imposes energetic and operational limitations~\cite{Liang2023,Zhou2021,Oyler2016,Yaacoub2021}.

Despite the increasing overlap of the distributions, the median cost exhibits a clear decrease for obstacle densities beyond the percolation threshold $\phi_c$. This decrease is most pronounced for smaller initial chaser populations, revealing an effect of cooperative interference among chasers. In strongly obstructed environments, the accessible space is confined to small clusters, and a large number of chasers leads to redundant trajectories and mutual blocking, thereby reducing overall efficiency. Under such conditions, a smaller population of chasers explores the limited accessible space more effectively, achieving a lower cost per captured escaper. This behavior parallels observations in biological group hunting and active-matter swarms, where optimal coordination rather than density maximization maximizes efficiency~\cite{Ohira2024,Hansen2023,Janosov2017,kamimura2010NJP}.

\subsection{Survival dynamics and Weibull statistics}

We now investigate the temporal evolution of the escaper population $N^{\rm E}(t)$ through survival analysis. This approach provides a complementary, time-resolved characterization of the pursuit dynamics, extending the global measures discussed previously, such as the trapping time and the cost. In particular, survival statistics allow us to resolve the coexistence of fast capture events and long-lived escapers arising from the geometric constraints identified by the BFS analysis.

\begin{figure}[h!]
    \centering
    \includegraphics[width=0.99\textwidth]{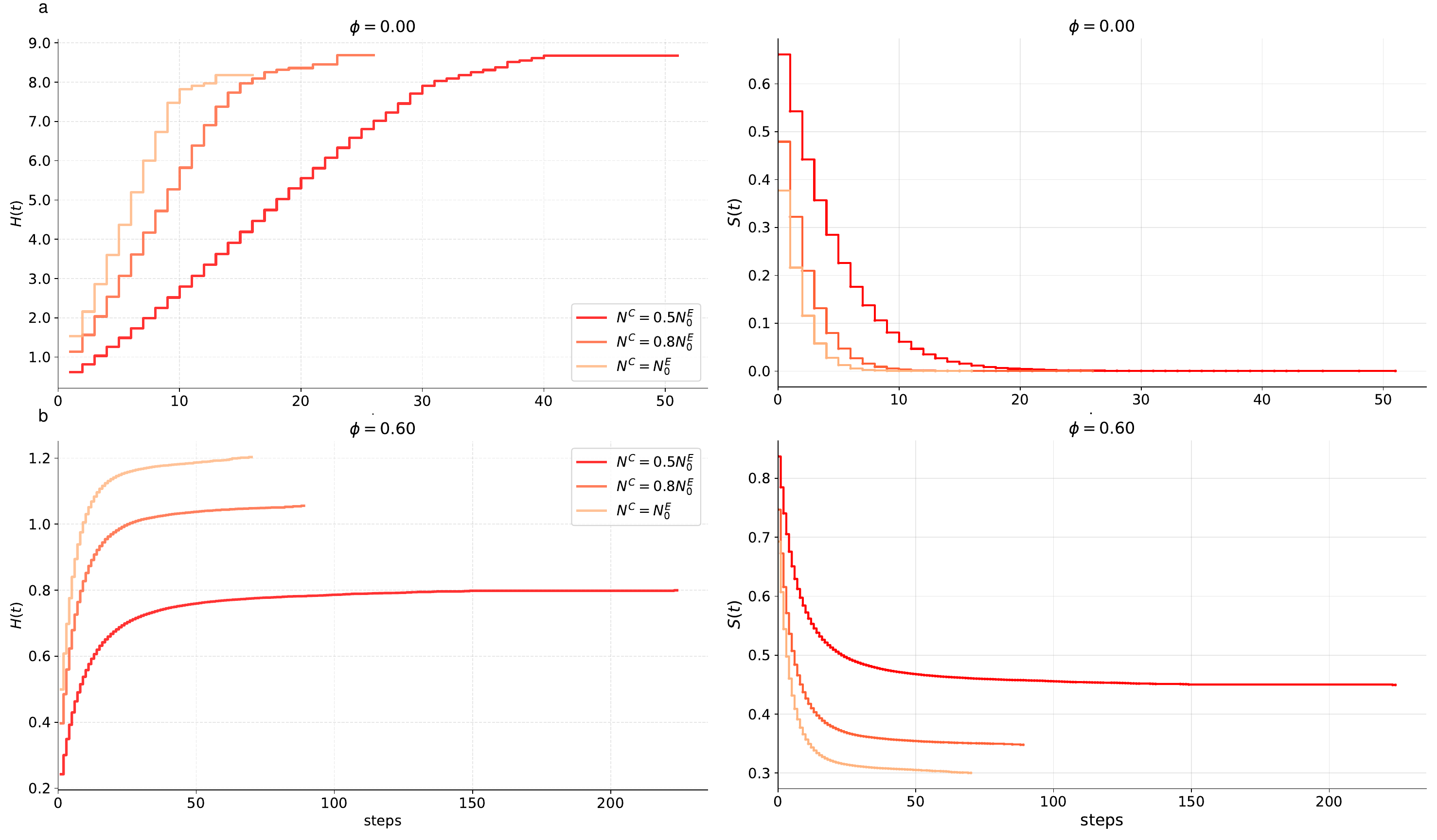}
    \caption{Cumulative hazard function $H(t)$ (left) and survival function $S(t)$ (right) for $\phi=0.00$ (a) and $\phi=\phi_c=0.60$ (b), shown for different initial chaser populations. Increasing obstacle density suppresses early hazards and enhances long-lived survival due to geometric fragmentation, while larger $N^{\rm C}$ increases the initial capture rate.}
    \label{panel_s_h}
\end{figure}

The survival function $S(t)$, defined in Eq.~\eqref{survival_function}, quantifies the probability that an escaper remains uncaptured up to time $t$, while the cumulative hazard function $H(t)$ captures the instantaneous risk of capture integrated over time. Figure~\ref{panel_s_h} shows $H(t)$ (left) and $S(t)$ (right) for two representative cases: an obstacle-free lattice ($\phi = 0.00$) and a strongly obstructed system near the percolation threshold ($\phi_c = 0.60$). Results are shown for different initial chaser populations $N^{\rm C}$.

A clear impact of spatial disorder on the hazard dynamics is observed. In the absence of obstacles, escapers experience a high initial hazard, reflecting frequent encounters in a fully connected environment. As $\phi$ increases, the cumulative hazard grows more slowly, indicating a reduced probability of encounters. This trend is consistent with the BFS results (Fig.~\ref{I.P}), which show that increasing obstacle density leads to spatial fragmentation and the isolation of escapers within disconnected clusters. As a consequence, geometric constraints suppress encounter rates and reshape the temporal structure of capture events.

For all obstacle densities, the initial hazard increases with the number of chasers, as expected from the higher encounter probability in denser pursuit configurations. This effect is most pronounced at early times, when escapers located near chasers are rapidly captured. The corresponding survival functions display a fast initial decay of $S(t)$, followed by a pronounced long-time tail. This two-stage behavior reflects the coexistence of cooperative pursuit at short times and slow geometric trapping at long times, in agreement with the nonmonotonic trends previously observed for the trapping time and the cost.

The decay of the escaper population does not follow a simple exponential law but instead exhibits stretched or compressed exponential behavior. This nontrivial temporal structure can be accurately described by a Weibull distribution, as previously reported in lattice-based chase--escape systems~\cite{vscepanovic2019PhysicaA},
\begin{equation}
N^{\rm E}(t) = A \exp\!\left[-\left(\frac{t}{\tau}\right)^{\beta}\right] + C,
\label{Weibull}
\end{equation}
with
\begin{equation}
C = N^{\rm E}(\infty), \qquad A = N^{\rm E}_0 - N^{\rm E}(\infty).
\label{S.E}
\end{equation}
Here, $\tau$ denotes the characteristic time of the decay, while the shape parameter $\beta$ quantifies deviations from exponential behavior and encodes the temporal heterogeneity of the capture process. The Weibull form naturally captures heterogeneous trapping processes characterized by broad distributions of waiting times, which are expected in fragmented and confined geometries.

\begin{figure}[h!]
    \centering
    \includegraphics[width=0.8\textwidth]{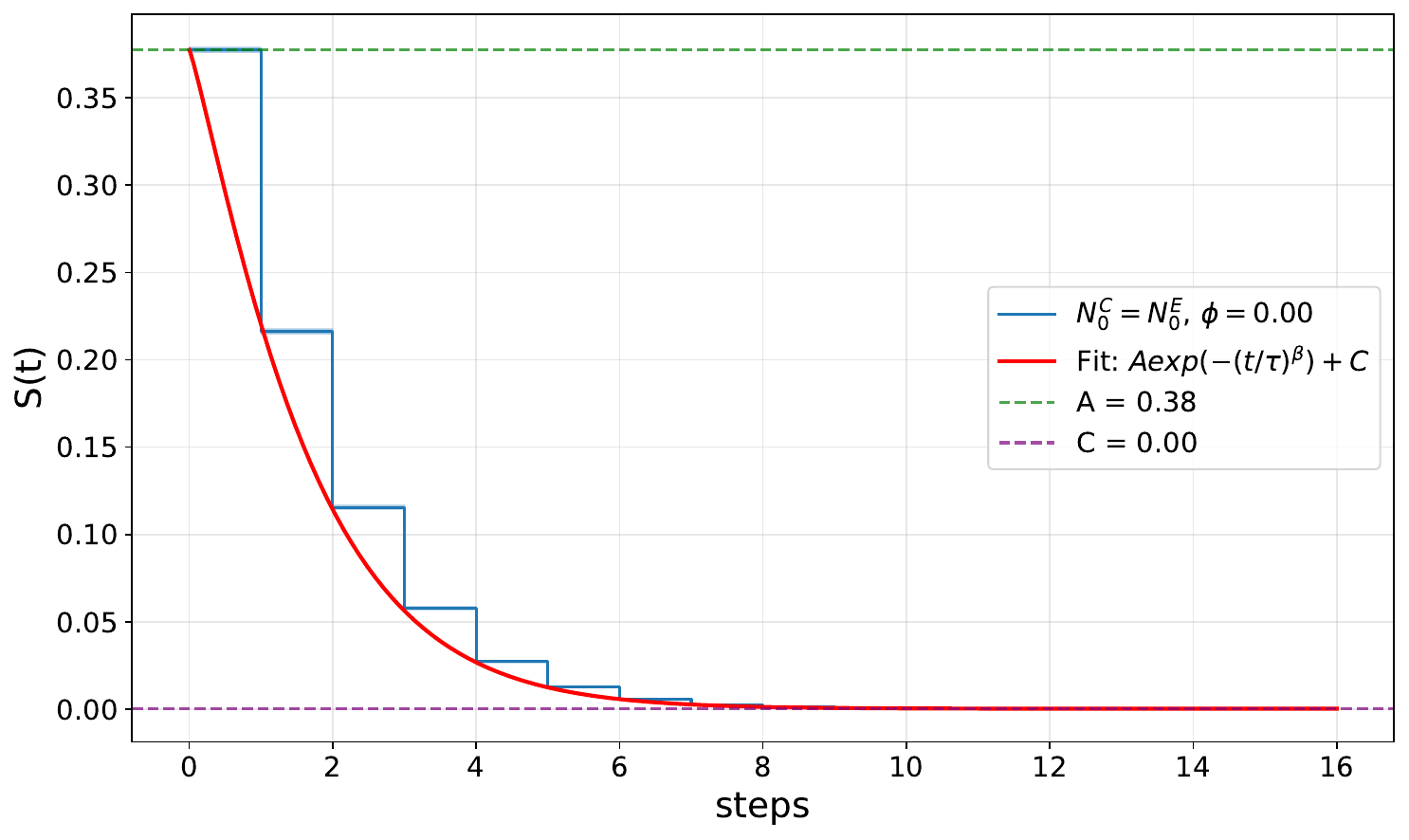}
    \caption{Weibull fit of the survival function $S(t)$ for $\phi=0$ and $N^{\rm C} = N^{\rm E}_0$. The fit accurately reproduces both the early-time decay and the long-time tail, validating the Weibull description of the survival dynamics.}
    \label{fit_S_0}
\end{figure}

The adequacy of the Weibull description is illustrated in Fig.~\ref{fit_S_0}, which shows representative fits for the obstacle-free case $\phi=0$ and $N^{\rm C} = N^{\rm E}_0$. The excellent agreement confirms that the Weibull form captures both the rapid initial decay associated with cooperative capture and the slow long-time tail arising from geometric trapping.

\begin{figure}[h!]
    \centering
    \includegraphics[width=0.8\textwidth]{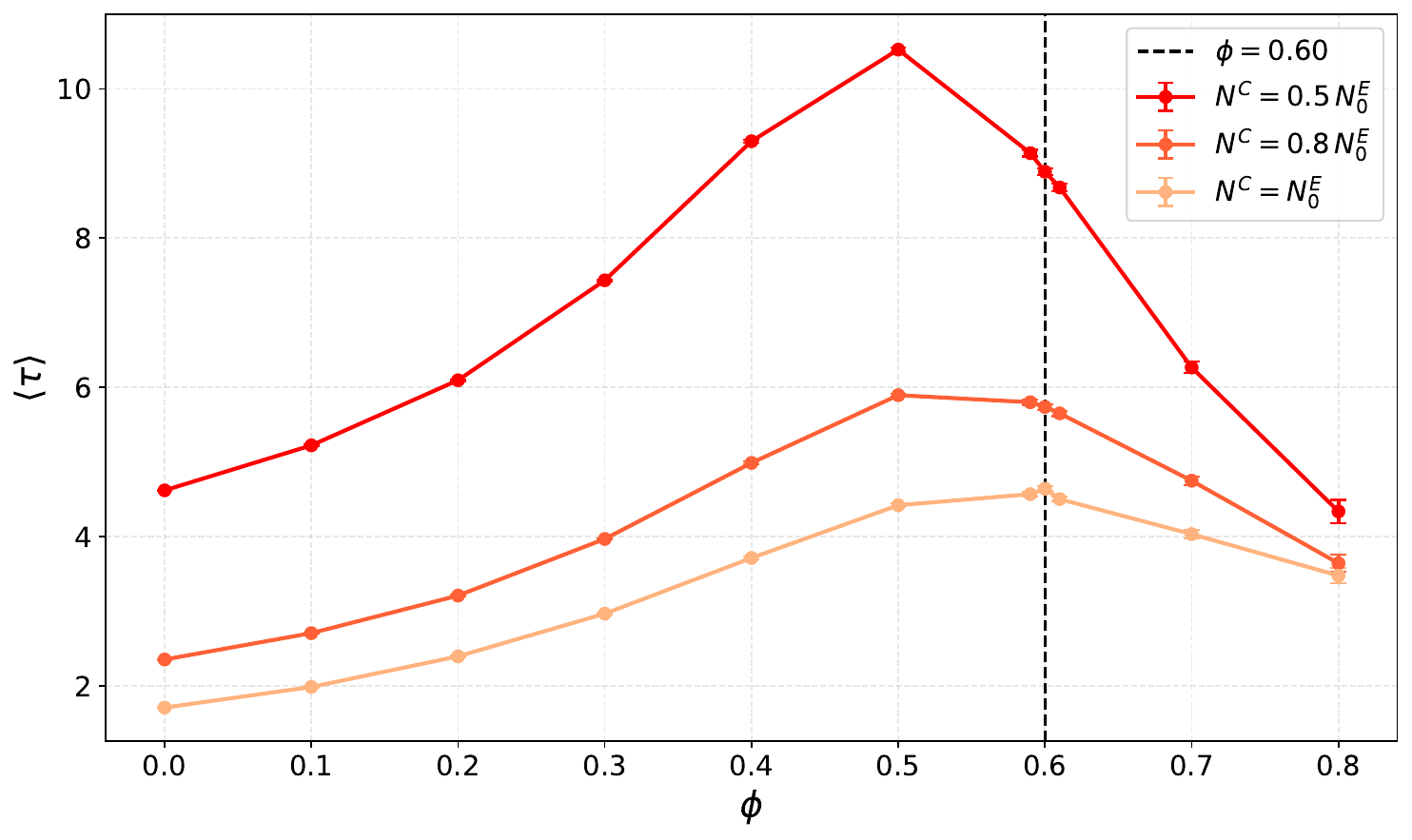}
    \caption{Characteristic time $\tau$ as a function of obstacle density $\phi$. The nonmonotonic behavior reflects the interplay between path elongation in connected lattices and geometric fragmentation near the percolation regime.}
    \label{tau_phi}
\end{figure}

We next examine how the characteristic time $\tau$ depends on the obstacle density $\phi$. Figure~\ref{tau_phi} shows that $\tau$ exhibits a nonmonotonic dependence on $\phi$, closely mirroring the behavior of the mean chaser--escaper distance obtained from the BFS analysis (Fig.~\ref{d_x_phi}). At low obstacle densities, the lattice remains well connected and encounters occur frequently, resulting in relatively short characteristic times. As $\phi$ increases, path elongation and local barriers hinder agent motion, leading to larger $\tau$ values. For smaller chaser populations ($N^{\rm C} = 0.5N^{\rm E}_0$), this effect is particularly pronounced, highlighting the enhanced sensitivity of sparse pursuit configurations to geometric disorder.

Beyond the connectivity breakdown, $\tau$ decreases sharply, reflecting the drastic reduction in the number of accessible escapers. In this regime, the remaining accessible clusters are small, and the escapers that remain reachable are captured over shorter absolute timescales. Similar geometry-controlled crossovers have been reported in percolation-based transport and trapping processes~\cite{Kumar2021,RevModPhys.88.045006,chepizhko13}.

\begin{figure}[h!]
    \centering
    \includegraphics[width=0.8\textwidth]{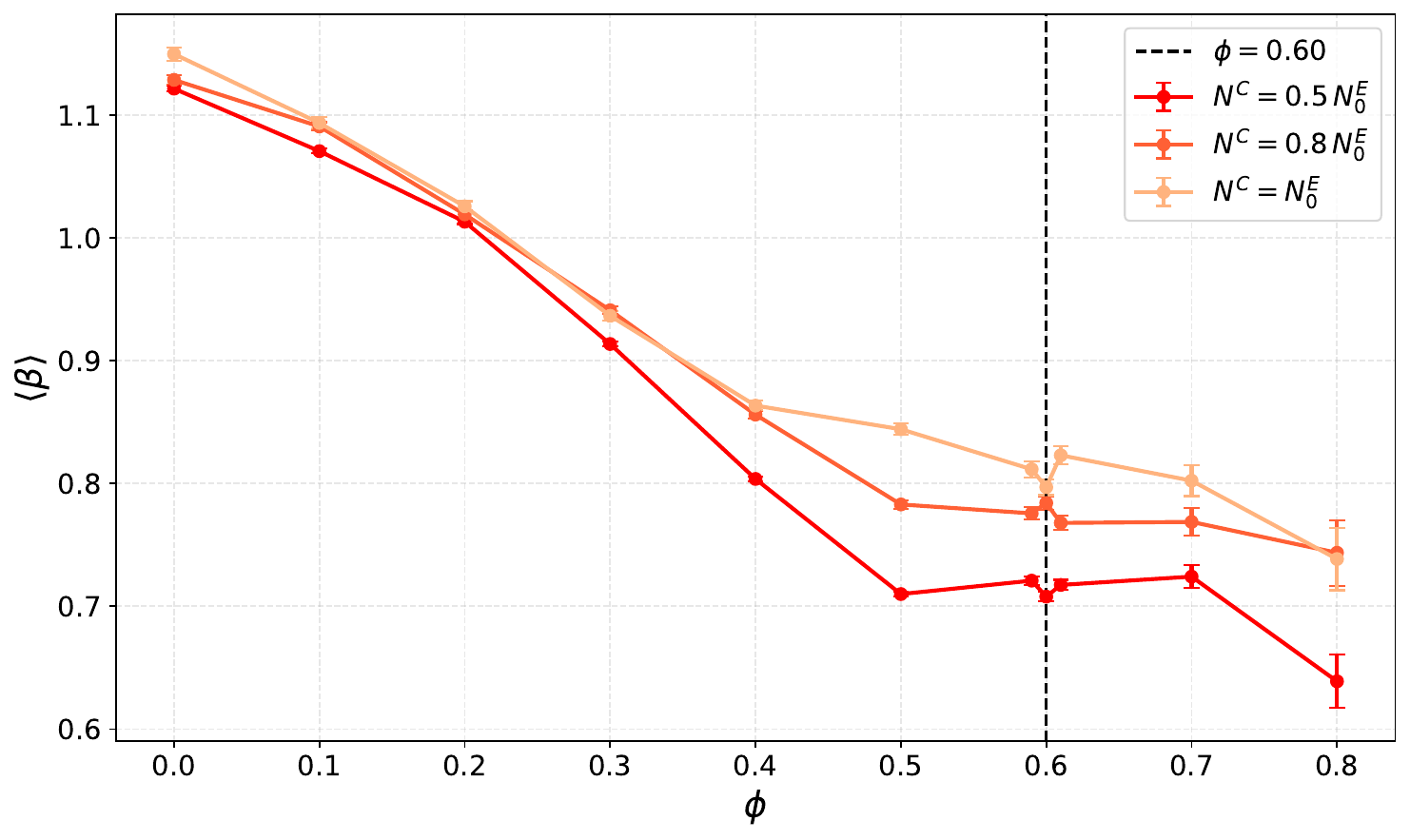}
    \caption{Shape parameter $\beta$ of the Weibull distribution as a function of obstacle density $\phi$. The crossover from $\beta>1$ to $\beta<1$ signals a transition from synchronized capture dynamics to heterogeneous, confinement-dominated survival.}
    \label{beta_phi}
\end{figure}

Further insight into the temporal structure of the survival dynamics is provided by the shape parameter $\beta$, shown in Fig.~\ref{beta_phi}. For $\phi \lesssim 0.30$, $\beta>1$, corresponding to a compressed-exponential regime in which captures are highly synchronized and occur predominantly at early times. As the obstacle density increases beyond this value, the system crosses over to a stretched-exponential regime with $\beta<1$, signaling the emergence of strong temporal heterogeneity and long-lived escapers. This transition reflects the onset of spatial confinement and broad distributions of trapping times, typical of reaction--diffusion and search processes in disordered media~\cite{Epstein2019,Bellomo2020}.

Interestingly, for $\phi>\phi_c$, the characteristic time $\tau$ decreases while $\beta$ remains below unity. This apparent contrast arises because, although the relative distribution of survival times remains broad, the absolute number of accessible escapers becomes small. Consequently, captures occur rapidly within isolated clusters, reducing $\tau$ without eliminating the heavy-tailed character of the survival statistics. Together, the trends in $\tau$ and $\beta$ identify two distinct dynamical regimes: a cooperative, diffusion-limited pursuit at low obstacle densities and a confinement-dominated trapping process at high obstacle densities.

The geometric and temporal heterogeneity revealed by the survival analysis is expected to leave clear signatures on transport properties, which we analyze next.

\subsection{Transport properties and mean-squared displacement}

\begin{figure}[h!]
    \centering
    \includegraphics[width=0.8\textwidth]{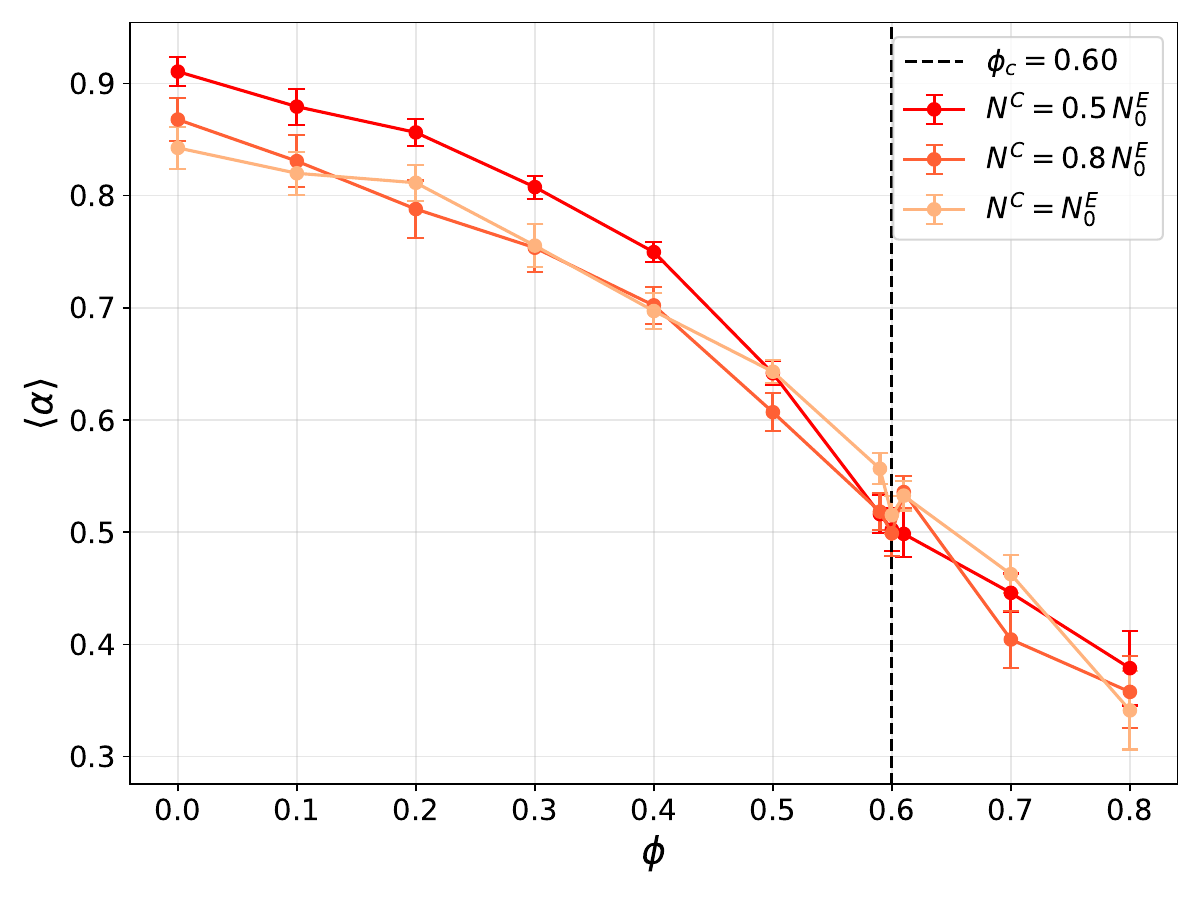}
    \caption{Mean-squared displacement exponent $\alpha$ as a function of obstacle density $\phi$. Subdiffusive behavior ($\alpha<1$) is observed even in obstacle-free systems due to collective interactions. With increasing $\phi$, geometric constraints dominate, leading to a convergence of $\alpha$ near the percolation threshold $\phi_c$, where transport becomes controlled by lattice connectivity.}
    \label{alpha_phi}
\end{figure}

We conclude the analysis by examining the transport properties of the system through the mean-squared displacement (MSD) of the chasers. Figure~\ref{alpha_phi} shows the dependence of the MSD scaling exponent $\alpha$ on the obstacle density $\phi$. The exponent $\alpha$ provides a direct measure of the effective transport regime, distinguishing normal diffusion from subdiffusive motion induced by interactions and geometric constraints.

Even in the absence of obstacles ($\phi=0$), the system exhibits subdiffusive behavior ($\alpha<1$). This effect originates from the high local density of agents and the resulting frequent encounters between chasers and escapers, which generate correlated motion, transient caging, and repeated reorientations. Although the lattice is fully connected in this regime, collective interactions alone are sufficient to suppress normal diffusion.

At low obstacle densities, agents can still explore large portions of the lattice, leading to relatively higher values of $\alpha$ compared to more obstructed configurations. As $\phi$ increases, however, geometric barriers progressively restrict agent motion. Path elongation, dead ends, and confinement within narrow corridors reduce mobility and enhance temporal correlations, driving the system deeper into the subdiffusive regime. This trend is consistent with the increase of the characteristic time $\tau$ and the emergence of stretched-exponential survival statistics discussed in the previous sections.

A key observation is the convergence of all $\alpha$ curves near the percolation threshold $\phi_c$. In this regime, the transport properties become largely independent of the number of chasers, indicating a transition to a geometry-dominated dynamical regime. Once the accessible space fragments into isolated clusters, the connectivity of these clusters dictates transport, overwhelming the influence of agent density and local interaction rules. This convergence mirrors the collapse observed for the trapping time, cost, and survival parameters, reinforcing the central role of percolation in controlling the system dynamics.

Taken together, the consistent behavior of the MSD exponent $\alpha$, the characteristic time $\tau$, and the Weibull shape parameter $\beta$ demonstrates that chase--escape dynamics in disordered lattices is governed by a competition between cooperative motion and geometric confinement. At low obstacle densities, collective pursuit and frequent interactions dominate, while at high $\phi$ transport becomes strongly subdiffusive and confined within isolated domains. The convergence of transport and temporal observables at $\phi_c$ highlights the universality of the geometric transition, linking spatial accessibility, temporal persistence, and collective behavior within a unified nonequilibrium framework~\cite{Bowick22,teVrugt2025,Kumar2021,RevModPhys.88.045006}.

\section{Conclusions}
We studied a minimal chase--and--escape model on a two-dimensional lattice with randomly distributed static obstacles, focusing on how geometric disorder controls collective pursuit dynamics. By combining geometric accessibility analysis, dynamical observables, survival statistics, and transport properties, we established a direct link between lattice connectivity and pursuit efficiency.

A Breadth-First Search (BFS) analysis showed that obstacle-induced fragmentation leads to a progressive loss of accessibility prior to the percolation threshold in the thermodynamic limit, defining the effective initial conditions for the dynamics. This geometric preconditioning shapes all subsequent observables. The trapping time and capture cost exhibit a nonmonotonic dependence on obstacle density, reflecting a competition between path elongation in connected lattices and geometric confinement near percolation. In highly obstructed environments, efficiency becomes largely independent of the number of chasers, indicating a geometry-dominated regime.

Survival analysis revealed that the decay of the escaper population follows a Weibull form, with characteristic time and shape parameters displaying clear crossovers as a function of obstacle density. These temporal signatures reflect the coexistence of cooperative capture at short times and confinement-dominated trapping at long times. Transport properties, quantified through the mean-squared displacement exponent, further reinforce this picture: subdiffusive motion arises from collective interactions at low obstacle densities and converges to a geometry-controlled regime near percolation.

Overall, our results demonstrate that chase--and--escape dynamics in disordered environments are governed by a geometry-driven crossover, where percolation and connectivity act as unifying control parameters for spatial, temporal, and collective behavior. This minimal framework provides insight into cooperative search processes in crowded and heterogeneous systems.

\section*{Acknowledgements}

Without public funding this research would be impossible. J.R.B. acknowledges financial support from Brazilian National Council for Scientific and Technological Development (CNPq), grant numbers 405479/2023-9, 441728/2023-5, and 304958/2022-0A, from the Research Support Foundation of the State of Rio Grande do Sul (FAPERGS), grant number 25/2551-0002609-1 and from the Coordination for the Improvement of Higher Education Personnel (CAPES, Brazil) and the Alexander von Humboldt Foundation for financial support through a research fellowship. RGR acknowledges support from CAPES (Finance Code 001) and CNPq, grant number 441728/2023-5.

\bibliography{apssamp}

\begin{thebibliography}{43}%
\makeatletter
\providecommand \@ifxundefined [1]{%
 \@ifx{#1\undefined}
}%
\providecommand \@ifnum [1]{%
 \ifnum #1\expandafter \@firstoftwo
 \else \expandafter \@secondoftwo
 \fi
}%
\providecommand \@ifx [1]{%
 \ifx #1\expandafter \@firstoftwo
 \else \expandafter \@secondoftwo
 \fi
}%
\providecommand \natexlab [1]{#1}%
\providecommand \enquote  [1]{``#1''}%
\providecommand \bibnamefont  [1]{#1}%
\providecommand \bibfnamefont [1]{#1}%
\providecommand \citenamefont [1]{#1}%
\providecommand \href@noop [0]{\@secondoftwo}%
\providecommand \href [0]{\begingroup \@sanitize@url \@href}%
\providecommand \@href[1]{\@@startlink{#1}\@@href}%
\providecommand \@@href[1]{\endgroup#1\@@endlink}%
\providecommand \@sanitize@url [0]{\catcode `\\12\catcode `\$12\catcode `\&12\catcode `\#12\catcode `\^12\catcode `\_12\catcode `\%12\relax}%
\providecommand \@@startlink[1]{}%
\providecommand \@@endlink[0]{}%
\providecommand \url  [0]{\begingroup\@sanitize@url \@url }%
\providecommand \@url [1]{\endgroup\@href {#1}{\urlprefix }}%
\providecommand \urlprefix  [0]{URL }%
\providecommand \Eprint [0]{\href }%
\providecommand \doibase [0]{https://doi.org/}%
\providecommand \selectlanguage [0]{\@gobble}%
\providecommand \bibinfo  [0]{\@secondoftwo}%
\providecommand \bibfield  [0]{\@secondoftwo}%
\providecommand \translation [1]{[#1]}%
\providecommand \BibitemOpen [0]{}%
\providecommand \bibitemStop [0]{}%
\providecommand \bibitemNoStop [0]{.\EOS\space}%
\providecommand \EOS [0]{\spacefactor3000\relax}%
\providecommand \BibitemShut  [1]{\csname bibitem#1\endcsname}%
\let\auto@bib@innerbib\@empty
\bibitem [{\citenamefont {Kamimura}\ and\ \citenamefont {Ohira}(2010)}]{kamimura2010NJP}%
  \BibitemOpen
  \bibfield  {author} {\bibinfo {author} {\bibfnamefont {A.}~\bibnamefont {Kamimura}}\ and\ \bibinfo {author} {\bibfnamefont {T.}~\bibnamefont {Ohira}},\ }\bibfield  {title} {\bibinfo {title} {Group chase and escape},\ }\href@noop {} {\bibfield  {journal} {\bibinfo  {journal} {New Journal of Physics}\ }\textbf {\bibinfo {volume} {12}},\ \bibinfo {pages} {053013} (\bibinfo {year} {2010})}\BibitemShut {NoStop}%
\bibitem [{\citenamefont {Angelani}(2012)}]{Angelani2012}%
  \BibitemOpen
  \bibfield  {author} {\bibinfo {author} {\bibfnamefont {L.}~\bibnamefont {Angelani}},\ }\bibfield  {title} {\bibinfo {title} {Collective predation and escape strategies},\ }\bibfield  {journal} {\bibinfo  {journal} {Physical Review Letters}\ }\textbf {\bibinfo {volume} {109}},\ \href {https://doi.org/10.1103/physrevlett.109.118104} {10.1103/physrevlett.109.118104} (\bibinfo {year} {2012})\BibitemShut {NoStop}%
\bibitem [{\citenamefont {Vicsek}\ and\ \citenamefont {Zafeiris}(2012)}]{Vicsek2012}%
  \BibitemOpen
  \bibfield  {author} {\bibinfo {author} {\bibfnamefont {T.}~\bibnamefont {Vicsek}}\ and\ \bibinfo {author} {\bibfnamefont {A.}~\bibnamefont {Zafeiris}},\ }\bibfield  {title} {\bibinfo {title} {Collective motion},\ }\href {https://doi.org/10.1016/j.physrep.2012.03.004} {\bibfield  {journal} {\bibinfo  {journal} {Physics Reports}\ }\textbf {\bibinfo {volume} {517}},\ \bibinfo {pages} {71–140} (\bibinfo {year} {2012})}\BibitemShut {NoStop}%
\bibitem [{\citenamefont {Janosov}\ \emph {et~al.}(2017)\citenamefont {Janosov}, \citenamefont {Virágh}, \citenamefont {Vásárhelyi},\ and\ \citenamefont {Vicsek}}]{Janosov2017}%
  \BibitemOpen
  \bibfield  {author} {\bibinfo {author} {\bibfnamefont {M.}~\bibnamefont {Janosov}}, \bibinfo {author} {\bibfnamefont {C.}~\bibnamefont {Virágh}}, \bibinfo {author} {\bibfnamefont {G.}~\bibnamefont {Vásárhelyi}},\ and\ \bibinfo {author} {\bibfnamefont {T.}~\bibnamefont {Vicsek}},\ }\bibfield  {title} {\bibinfo {title} {Group chasing tactics: how to catch a faster prey},\ }\href {https://doi.org/10.1088/1367-2630/aa69e7} {\bibfield  {journal} {\bibinfo  {journal} {New Journal of Physics}\ }\textbf {\bibinfo {volume} {19}},\ \bibinfo {pages} {053003} (\bibinfo {year} {2017})}\BibitemShut {NoStop}%
\bibitem [{\citenamefont {Ohira}(2024)}]{Ohira2024}%
  \BibitemOpen
  \bibfield  {author} {\bibinfo {author} {\bibfnamefont {T.}~\bibnamefont {Ohira}},\ }\bibfield  {title} {\bibinfo {title} {Collective behaviors emerging from chases and escapes},\ }\href {https://doi.org/10.1007/s10015-023-00928-1} {\bibfield  {journal} {\bibinfo  {journal} {Artificial Life and Robotics}\ }\textbf {\bibinfo {volume} {29}},\ \bibinfo {pages} {1–11} (\bibinfo {year} {2024})}\BibitemShut {NoStop}%
\bibitem [{\citenamefont {Bechinger}\ \emph {et~al.}(2016)\citenamefont {Bechinger}, \citenamefont {Di~Leonardo}, \citenamefont {L\"owen}, \citenamefont {Reichhardt}, \citenamefont {Volpe},\ and\ \citenamefont {Volpe}}]{RevModPhys.88.045006}%
  \BibitemOpen
  \bibfield  {author} {\bibinfo {author} {\bibfnamefont {C.}~\bibnamefont {Bechinger}}, \bibinfo {author} {\bibfnamefont {R.}~\bibnamefont {Di~Leonardo}}, \bibinfo {author} {\bibfnamefont {H.}~\bibnamefont {L\"owen}}, \bibinfo {author} {\bibfnamefont {C.}~\bibnamefont {Reichhardt}}, \bibinfo {author} {\bibfnamefont {G.}~\bibnamefont {Volpe}},\ and\ \bibinfo {author} {\bibfnamefont {G.}~\bibnamefont {Volpe}},\ }\bibfield  {title} {\bibinfo {title} {Active particles in complex and crowded environments},\ }\href {https://doi.org/10.1103/RevModPhys.88.045006} {\bibfield  {journal} {\bibinfo  {journal} {Rev. Mod. Phys.}\ }\textbf {\bibinfo {volume} {88}},\ \bibinfo {pages} {045006} (\bibinfo {year} {2016})}\BibitemShut {NoStop}%
\bibitem [{\citenamefont {Bowick}\ \emph {et~al.}(2022)\citenamefont {Bowick}, \citenamefont {Fakhri}, \citenamefont {Marchetti},\ and\ \citenamefont {Ramaswamy}}]{Bowick22}%
  \BibitemOpen
  \bibfield  {author} {\bibinfo {author} {\bibfnamefont {M.~J.}\ \bibnamefont {Bowick}}, \bibinfo {author} {\bibfnamefont {N.}~\bibnamefont {Fakhri}}, \bibinfo {author} {\bibfnamefont {M.~C.}\ \bibnamefont {Marchetti}},\ and\ \bibinfo {author} {\bibfnamefont {S.}~\bibnamefont {Ramaswamy}},\ }\bibfield  {title} {\bibinfo {title} {Symmetry, thermodynamics, and topology in active matter},\ }\href {https://doi.org/10.1103/PhysRevX.12.010501} {\bibfield  {journal} {\bibinfo  {journal} {Phys. Rev. X}\ }\textbf {\bibinfo {volume} {12}},\ \bibinfo {pages} {010501} (\bibinfo {year} {2022})}\BibitemShut {NoStop}%
\bibitem [{\citenamefont {Epstein}\ \emph {et~al.}(2019)\citenamefont {Epstein}, \citenamefont {Klymko},\ and\ \citenamefont {Mandadapu}}]{Epstein2019}%
  \BibitemOpen
  \bibfield  {author} {\bibinfo {author} {\bibfnamefont {J.~M.}\ \bibnamefont {Epstein}}, \bibinfo {author} {\bibfnamefont {K.}~\bibnamefont {Klymko}},\ and\ \bibinfo {author} {\bibfnamefont {K.~K.}\ \bibnamefont {Mandadapu}},\ }\bibfield  {title} {\bibinfo {title} {Statistical mechanics of transport processes in active fluids. ii. equations of hydrodynamics for active brownian particles},\ }\bibfield  {journal} {\bibinfo  {journal} {The Journal of Chemical Physics}\ }\textbf {\bibinfo {volume} {150}},\ \href {https://doi.org/10.1063/1.5054912} {10.1063/1.5054912} (\bibinfo {year} {2019})\BibitemShut {NoStop}%
\bibitem [{\citenamefont {te~Vrugt}\ and\ \citenamefont {Wittkowski}(2025)}]{teVrugt2025}%
  \BibitemOpen
  \bibfield  {author} {\bibinfo {author} {\bibfnamefont {M.}~\bibnamefont {te~Vrugt}}\ and\ \bibinfo {author} {\bibfnamefont {R.}~\bibnamefont {Wittkowski}},\ }\bibfield  {title} {\bibinfo {title} {Metareview: a survey of active matter reviews},\ }\bibfield  {journal} {\bibinfo  {journal} {The European Physical Journal E}\ }\textbf {\bibinfo {volume} {48}},\ \href {https://doi.org/10.1140/epje/s10189-024-00466-z} {10.1140/epje/s10189-024-00466-z} (\bibinfo {year} {2025})\BibitemShut {NoStop}%
\bibitem [{\citenamefont {Evans}(1993)}]{evans1993RMP}%
  \BibitemOpen
  \bibfield  {author} {\bibinfo {author} {\bibfnamefont {J.~W.}\ \bibnamefont {Evans}},\ }\bibfield  {title} {\bibinfo {title} {Random and cooperative sequential adsorption},\ }\href@noop {} {\bibfield  {journal} {\bibinfo  {journal} {Reviews of modern physics}\ }\textbf {\bibinfo {volume} {65}},\ \bibinfo {pages} {1281} (\bibinfo {year} {1993})}\BibitemShut {NoStop}%
\bibitem [{\citenamefont {Fowell}\ and\ \citenamefont {Kim}(2021)}]{Fowell2021}%
  \BibitemOpen
  \bibfield  {author} {\bibinfo {author} {\bibfnamefont {D.~J.}\ \bibnamefont {Fowell}}\ and\ \bibinfo {author} {\bibfnamefont {M.}~\bibnamefont {Kim}},\ }\bibfield  {title} {\bibinfo {title} {The spatio-temporal control of effector t cell migration},\ }\href {https://doi.org/10.1038/s41577-021-00507-0} {\bibfield  {journal} {\bibinfo  {journal} {Nature Reviews Immunology}\ }\textbf {\bibinfo {volume} {21}},\ \bibinfo {pages} {582–596} (\bibinfo {year} {2021})}\BibitemShut {NoStop}%
\bibitem [{\citenamefont {Schienstock}\ and\ \citenamefont {Mueller}(2021)}]{Schienstock2021}%
  \BibitemOpen
  \bibfield  {author} {\bibinfo {author} {\bibfnamefont {D.}~\bibnamefont {Schienstock}}\ and\ \bibinfo {author} {\bibfnamefont {S.~N.}\ \bibnamefont {Mueller}},\ }\bibfield  {title} {\bibinfo {title} {Moving beyond velocity: Opportunities and challenges to quantify immune cell behavior*},\ }\href {https://doi.org/10.1111/imr.13038} {\bibfield  {journal} {\bibinfo  {journal} {Immunological Reviews}\ }\textbf {\bibinfo {volume} {306}},\ \bibinfo {pages} {123–136} (\bibinfo {year} {2021})}\BibitemShut {NoStop}%
\bibitem [{\citenamefont {Moreira-Soares}\ \emph {et~al.}(2020)\citenamefont {Moreira-Soares}, \citenamefont {Cunha}, \citenamefont {Bordin},\ and\ \citenamefont {Travasso}}]{MoreiraSoares2020}%
  \BibitemOpen
  \bibfield  {author} {\bibinfo {author} {\bibfnamefont {M.}~\bibnamefont {Moreira-Soares}}, \bibinfo {author} {\bibfnamefont {S.~P.}\ \bibnamefont {Cunha}}, \bibinfo {author} {\bibfnamefont {J.~R.}\ \bibnamefont {Bordin}},\ and\ \bibinfo {author} {\bibfnamefont {R.~D.~M.}\ \bibnamefont {Travasso}},\ }\bibfield  {title} {\bibinfo {title} {Adhesion modulates cell morphology and migration within dense fibrous networks},\ }\href {https://doi.org/10.1088/1361-648x/ab7c17} {\bibfield  {journal} {\bibinfo  {journal} {Journal of Physics: Condensed Matter}\ }\textbf {\bibinfo {volume} {32}},\ \bibinfo {pages} {314001} (\bibinfo {year} {2020})}\BibitemShut {NoStop}%
\bibitem [{\citenamefont {Melo}\ \emph {et~al.}(2023)\citenamefont {Melo}, \citenamefont {Guerrero}, \citenamefont {Moreira~Soares}, \citenamefont {Bordin}, \citenamefont {Carneiro}, \citenamefont {Carneiro}, \citenamefont {Dias}, \citenamefont {Carvalho}, \citenamefont {Figueiredo}, \citenamefont {Seruca},\ and\ \citenamefont {Travasso}}]{Melo2023}%
  \BibitemOpen
  \bibfield  {author} {\bibinfo {author} {\bibfnamefont {S.}~\bibnamefont {Melo}}, \bibinfo {author} {\bibfnamefont {P.}~\bibnamefont {Guerrero}}, \bibinfo {author} {\bibfnamefont {M.}~\bibnamefont {Moreira~Soares}}, \bibinfo {author} {\bibfnamefont {J.~R.}\ \bibnamefont {Bordin}}, \bibinfo {author} {\bibfnamefont {F.}~\bibnamefont {Carneiro}}, \bibinfo {author} {\bibfnamefont {P.}~\bibnamefont {Carneiro}}, \bibinfo {author} {\bibfnamefont {M.~B.}\ \bibnamefont {Dias}}, \bibinfo {author} {\bibfnamefont {J.}~\bibnamefont {Carvalho}}, \bibinfo {author} {\bibfnamefont {J.}~\bibnamefont {Figueiredo}}, \bibinfo {author} {\bibfnamefont {R.}~\bibnamefont {Seruca}},\ and\ \bibinfo {author} {\bibfnamefont {R.~D.~M.}\ \bibnamefont {Travasso}},\ }\bibfield  {title} {\bibinfo {title} {The ecm and tissue architecture are major determinants of early invasion mediated by e-cadherin dysfunction},\ }\bibfield  {journal} {\bibinfo  {journal} {Communications Biology}\ }\textbf {\bibinfo {volume} {6}},\ \href
  {https://doi.org/10.1038/s42003-023-05482-x} {10.1038/s42003-023-05482-x} (\bibinfo {year} {2023})\BibitemShut {NoStop}%
\bibitem [{\citenamefont {GORINI}\ \emph {et~al.}(2011)\citenamefont {GORINI}, \citenamefont {LINNELL}, \citenamefont {MAY}, \citenamefont {PANZACCHI}, \citenamefont {BOITANI}, \citenamefont {ODDEN},\ and\ \citenamefont {NILSEN}}]{Gorini11}%
  \BibitemOpen
  \bibfield  {author} {\bibinfo {author} {\bibfnamefont {L.}~\bibnamefont {GORINI}}, \bibinfo {author} {\bibfnamefont {J.~D.~C.}\ \bibnamefont {LINNELL}}, \bibinfo {author} {\bibfnamefont {R.}~\bibnamefont {MAY}}, \bibinfo {author} {\bibfnamefont {M.}~\bibnamefont {PANZACCHI}}, \bibinfo {author} {\bibfnamefont {L.}~\bibnamefont {BOITANI}}, \bibinfo {author} {\bibfnamefont {M.}~\bibnamefont {ODDEN}},\ and\ \bibinfo {author} {\bibfnamefont {E.~B.}\ \bibnamefont {NILSEN}},\ }\bibfield  {title} {\bibinfo {title} {Habitat heterogeneity and mammalian predator–prey interactions},\ }\href {https://doi.org/10.1111/j.1365-2907.2011.00189.x} {\bibfield  {journal} {\bibinfo  {journal} {Mammal Review}\ }\textbf {\bibinfo {volume} {42}},\ \bibinfo {pages} {55–77} (\bibinfo {year} {2011})}\BibitemShut {NoStop}%
\bibitem [{\citenamefont {Cozzi}\ \emph {et~al.}(2013)\citenamefont {Cozzi}, \citenamefont {Broekhuis}, \citenamefont {McNutt},\ and\ \citenamefont {Schmid}}]{Cozzi2013}%
  \BibitemOpen
  \bibfield  {author} {\bibinfo {author} {\bibfnamefont {G.}~\bibnamefont {Cozzi}}, \bibinfo {author} {\bibfnamefont {F.}~\bibnamefont {Broekhuis}}, \bibinfo {author} {\bibfnamefont {J.~W.}\ \bibnamefont {McNutt}},\ and\ \bibinfo {author} {\bibfnamefont {B.}~\bibnamefont {Schmid}},\ }\bibfield  {title} {\bibinfo {title} {Comparison of the effects of artificial and natural barriers on large african carnivores: Implications for interspecific relationships and connectivity},\ }\href {https://doi.org/10.1111/1365-2656.12039} {\bibfield  {journal} {\bibinfo  {journal} {Journal of Animal Ecology}\ }\textbf {\bibinfo {volume} {82}},\ \bibinfo {pages} {707–715} (\bibinfo {year} {2013})}\BibitemShut {NoStop}%
\bibitem [{\citenamefont {Oyler}\ \emph {et~al.}(2016)\citenamefont {Oyler}, \citenamefont {Kabamba},\ and\ \citenamefont {Girard}}]{Oyler2016}%
  \BibitemOpen
  \bibfield  {author} {\bibinfo {author} {\bibfnamefont {D.~W.}\ \bibnamefont {Oyler}}, \bibinfo {author} {\bibfnamefont {P.~T.}\ \bibnamefont {Kabamba}},\ and\ \bibinfo {author} {\bibfnamefont {A.~R.}\ \bibnamefont {Girard}},\ }\bibfield  {title} {\bibinfo {title} {Pursuit–evasion games in the presence of obstacles},\ }\href {https://doi.org/10.1016/j.automatica.2015.11.018} {\bibfield  {journal} {\bibinfo  {journal} {Automatica}\ }\textbf {\bibinfo {volume} {65}},\ \bibinfo {pages} {1–11} (\bibinfo {year} {2016})}\BibitemShut {NoStop}%
\bibitem [{\citenamefont {Liang}\ \emph {et~al.}(2023)\citenamefont {Liang}, \citenamefont {Zhou}, \citenamefont {Jiang}, \citenamefont {Meng},\ and\ \citenamefont {Xiu}}]{Liang2023}%
  \BibitemOpen
  \bibfield  {author} {\bibinfo {author} {\bibfnamefont {X.}~\bibnamefont {Liang}}, \bibinfo {author} {\bibfnamefont {B.}~\bibnamefont {Zhou}}, \bibinfo {author} {\bibfnamefont {L.}~\bibnamefont {Jiang}}, \bibinfo {author} {\bibfnamefont {G.}~\bibnamefont {Meng}},\ and\ \bibinfo {author} {\bibfnamefont {Y.}~\bibnamefont {Xiu}},\ }\bibfield  {title} {\bibinfo {title} {Collaborative pursuit-evasion game of multi-uavs based on apollonius circle in the environment with obstacle},\ }\bibfield  {journal} {\bibinfo  {journal} {Connection Science}\ }\textbf {\bibinfo {volume} {35}},\ \href {https://doi.org/10.1080/09540091.2023.2168253} {10.1080/09540091.2023.2168253} (\bibinfo {year} {2023})\BibitemShut {NoStop}%
\bibitem [{\citenamefont {Katona}\ \emph {et~al.}(2024)\citenamefont {Katona}, \citenamefont {Neamah},\ and\ \citenamefont {Korondi}}]{Katona2024}%
  \BibitemOpen
  \bibfield  {author} {\bibinfo {author} {\bibfnamefont {K.}~\bibnamefont {Katona}}, \bibinfo {author} {\bibfnamefont {H.~A.}\ \bibnamefont {Neamah}},\ and\ \bibinfo {author} {\bibfnamefont {P.}~\bibnamefont {Korondi}},\ }\bibfield  {title} {\bibinfo {title} {Obstacle avoidance and path planning methods for autonomous navigation of mobile robot},\ }\href {https://doi.org/10.3390/s24113573} {\bibfield  {journal} {\bibinfo  {journal} {Sensors}\ }\textbf {\bibinfo {volume} {24}},\ \bibinfo {pages} {3573} (\bibinfo {year} {2024})}\BibitemShut {NoStop}%
\bibitem [{\citenamefont {Zhou}\ \emph {et~al.}(2021)\citenamefont {Zhou}, \citenamefont {Chen}, \citenamefont {He},\ and\ \citenamefont {Bian}}]{Zhou2021}%
  \BibitemOpen
  \bibfield  {author} {\bibinfo {author} {\bibfnamefont {Y.}~\bibnamefont {Zhou}}, \bibinfo {author} {\bibfnamefont {A.}~\bibnamefont {Chen}}, \bibinfo {author} {\bibfnamefont {X.}~\bibnamefont {He}},\ and\ \bibinfo {author} {\bibfnamefont {X.}~\bibnamefont {Bian}},\ }\bibfield  {title} {\bibinfo {title} {Multi-target coordinated search algorithm for swarm robotics considering practical constraints},\ }\bibfield  {journal} {\bibinfo  {journal} {Frontiers in Neurorobotics}\ }\textbf {\bibinfo {volume} {15}},\ \href {https://doi.org/10.3389/fnbot.2021.753052} {10.3389/fnbot.2021.753052} (\bibinfo {year} {2021})\BibitemShut {NoStop}%
\bibitem [{\citenamefont {Yaacoub}\ \emph {et~al.}(2021)\citenamefont {Yaacoub}, \citenamefont {Noura}, \citenamefont {Salman},\ and\ \citenamefont {Chehab}}]{Yaacoub2021}%
  \BibitemOpen
  \bibfield  {author} {\bibinfo {author} {\bibfnamefont {J.-P.~A.}\ \bibnamefont {Yaacoub}}, \bibinfo {author} {\bibfnamefont {H.~N.}\ \bibnamefont {Noura}}, \bibinfo {author} {\bibfnamefont {O.}~\bibnamefont {Salman}},\ and\ \bibinfo {author} {\bibfnamefont {A.}~\bibnamefont {Chehab}},\ }\bibfield  {title} {\bibinfo {title} {Robotics cyber security: vulnerabilities, attacks, countermeasures, and recommendations},\ }\href {https://doi.org/10.1007/s10207-021-00545-8} {\bibfield  {journal} {\bibinfo  {journal} {International Journal of Information Security}\ }\textbf {\bibinfo {volume} {21}},\ \bibinfo {pages} {115–158} (\bibinfo {year} {2021})}\BibitemShut {NoStop}%
\bibitem [{\citenamefont {Alqudsi}\ and\ \citenamefont {Makaraci}(2025)}]{Alqudsi2025}%
  \BibitemOpen
  \bibfield  {author} {\bibinfo {author} {\bibfnamefont {Y.}~\bibnamefont {Alqudsi}}\ and\ \bibinfo {author} {\bibfnamefont {M.}~\bibnamefont {Makaraci}},\ }\bibfield  {title} {\bibinfo {title} {Uav swarms: research, challenges, and future directions},\ }\bibfield  {journal} {\bibinfo  {journal} {Journal of Engineering and Applied Science}\ }\textbf {\bibinfo {volume} {72}},\ \href {https://doi.org/10.1186/s44147-025-00582-3} {10.1186/s44147-025-00582-3} (\bibinfo {year} {2025})\BibitemShut {NoStop}%
\bibitem [{\citenamefont {Helbing}\ \emph {et~al.}(2005)\citenamefont {Helbing}, \citenamefont {Buzna}, \citenamefont {Johansson},\ and\ \citenamefont {Werner}}]{helbing2005TS}%
  \BibitemOpen
  \bibfield  {author} {\bibinfo {author} {\bibfnamefont {D.}~\bibnamefont {Helbing}}, \bibinfo {author} {\bibfnamefont {L.}~\bibnamefont {Buzna}}, \bibinfo {author} {\bibfnamefont {A.}~\bibnamefont {Johansson}},\ and\ \bibinfo {author} {\bibfnamefont {T.}~\bibnamefont {Werner}},\ }\bibfield  {title} {\bibinfo {title} {Self-organized pedestrian crowd dynamics: Experiments, simulations, and design solutions},\ }\href@noop {} {\bibfield  {journal} {\bibinfo  {journal} {Transportation science}\ }\textbf {\bibinfo {volume} {39}},\ \bibinfo {pages} {1} (\bibinfo {year} {2005})}\BibitemShut {NoStop}%
\bibitem [{\citenamefont {Zhang}\ \emph {et~al.}(2021)\citenamefont {Zhang}, \citenamefont {Huang}, \citenamefont {Ji}, \citenamefont {Liu},\ and\ \citenamefont {Tang}}]{zhang2021PRA}%
  \BibitemOpen
  \bibfield  {author} {\bibinfo {author} {\bibfnamefont {D.}~\bibnamefont {Zhang}}, \bibinfo {author} {\bibfnamefont {G.}~\bibnamefont {Huang}}, \bibinfo {author} {\bibfnamefont {C.}~\bibnamefont {Ji}}, \bibinfo {author} {\bibfnamefont {H.}~\bibnamefont {Liu}},\ and\ \bibinfo {author} {\bibfnamefont {Y.}~\bibnamefont {Tang}},\ }\bibfield  {title} {\bibinfo {title} {Pedestrian evacuation modeling and simulation in multi-exit scenarios},\ }\href@noop {} {\bibfield  {journal} {\bibinfo  {journal} {Physica A: Statistical Mechanics and its Applications}\ }\textbf {\bibinfo {volume} {582}},\ \bibinfo {pages} {126272} (\bibinfo {year} {2021})}\BibitemShut {NoStop}%
\bibitem [{\citenamefont {Wang}\ \emph {et~al.}(2025)\citenamefont {Wang}, \citenamefont {Ge},\ and\ \citenamefont {Comber}}]{wang2025JCSS}%
  \BibitemOpen
  \bibfield  {author} {\bibinfo {author} {\bibfnamefont {Y.}~\bibnamefont {Wang}}, \bibinfo {author} {\bibfnamefont {J.}~\bibnamefont {Ge}},\ and\ \bibinfo {author} {\bibfnamefont {A.}~\bibnamefont {Comber}},\ }\bibfield  {title} {\bibinfo {title} {Modelling emergent pedestrian evacuation behaviors from intelligent, game-playing agents},\ }\href@noop {} {\bibfield  {journal} {\bibinfo  {journal} {Journal of Computational Social Science}\ }\textbf {\bibinfo {volume} {8}},\ \bibinfo {pages} {49} (\bibinfo {year} {2025})}\BibitemShut {NoStop}%
\bibitem [{\citenamefont {Chepizhko}\ and\ \citenamefont {Peruani}(2013)}]{chepizhko13}%
  \BibitemOpen
  \bibfield  {author} {\bibinfo {author} {\bibfnamefont {O.}~\bibnamefont {Chepizhko}}\ and\ \bibinfo {author} {\bibfnamefont {F.}~\bibnamefont {Peruani}},\ }\bibfield  {title} {\bibinfo {title} {Diffusion, subdiffusion, and trapping of active particles in heterogeneous media},\ }\href {https://doi.org/10.1103/PhysRevLett.111.160604} {\bibfield  {journal} {\bibinfo  {journal} {Phys. Rev. Lett.}\ }\textbf {\bibinfo {volume} {111}},\ \bibinfo {pages} {160604} (\bibinfo {year} {2013})}\BibitemShut {NoStop}%
\bibitem [{\citenamefont {Bellomo}\ \emph {et~al.}(2020)\citenamefont {Bellomo}, \citenamefont {Brezzi},\ and\ \citenamefont {Soler}}]{Bellomo2020}%
  \BibitemOpen
  \bibfield  {author} {\bibinfo {author} {\bibfnamefont {N.}~\bibnamefont {Bellomo}}, \bibinfo {author} {\bibfnamefont {F.}~\bibnamefont {Brezzi}},\ and\ \bibinfo {author} {\bibfnamefont {J.}~\bibnamefont {Soler}},\ }\bibfield  {title} {\bibinfo {title} {Active particles methods and challenges in behavioral systems},\ }\href {https://doi.org/10.1142/s0218202520020017} {\bibfield  {journal} {\bibinfo  {journal} {Mathematical Models and Methods in Applied Sciences}\ }\textbf {\bibinfo {volume} {30}},\ \bibinfo {pages} {653–658} (\bibinfo {year} {2020})}\BibitemShut {NoStop}%
\bibitem [{\citenamefont {Kumar}\ \emph {et~al.}(2021)\citenamefont {Kumar}, \citenamefont {Grassberger},\ and\ \citenamefont {Dhar}}]{Kumar2021}%
  \BibitemOpen
  \bibfield  {author} {\bibinfo {author} {\bibfnamefont {A.}~\bibnamefont {Kumar}}, \bibinfo {author} {\bibfnamefont {P.}~\bibnamefont {Grassberger}},\ and\ \bibinfo {author} {\bibfnamefont {D.}~\bibnamefont {Dhar}},\ }\bibfield  {title} {\bibinfo {title} {Chase-escape percolation on the 2d square lattice},\ }\href {https://doi.org/10.1016/j.physa.2021.126072} {\bibfield  {journal} {\bibinfo  {journal} {Physica A: Statistical Mechanics and its Applications}\ }\textbf {\bibinfo {volume} {577}},\ \bibinfo {pages} {126072} (\bibinfo {year} {2021})}\BibitemShut {NoStop}%
\bibitem [{\citenamefont {Cornette}\ \emph {et~al.}(2003)\citenamefont {Cornette}, \citenamefont {Ramirez-Pastor},\ and\ \citenamefont {Nieto}}]{cornette2003Springer}%
  \BibitemOpen
  \bibfield  {author} {\bibinfo {author} {\bibfnamefont {V.}~\bibnamefont {Cornette}}, \bibinfo {author} {\bibfnamefont {A.~J.}\ \bibnamefont {Ramirez-Pastor}},\ and\ \bibinfo {author} {\bibfnamefont {F.}~\bibnamefont {Nieto}},\ }\bibfield  {title} {\bibinfo {title} {Percolation of polyatomic species on a square lattice},\ }\href@noop {} {\bibfield  {journal} {\bibinfo  {journal} {The European Physical Journal B-Condensed Matter and Complex Systems}\ }\textbf {\bibinfo {volume} {36}},\ \bibinfo {pages} {391} (\bibinfo {year} {2003})}\BibitemShut {NoStop}%
\bibitem [{\citenamefont {{\v{S}}{\'c}epanovi{\'c}}\ \emph {et~al.}(2019)\citenamefont {{\v{S}}{\'c}epanovi{\'c}}, \citenamefont {Kara{\v{c}}}, \citenamefont {Jak{\v{s}}i{\'c}}, \citenamefont {Budinski-Petkovi{\'c}},\ and\ \citenamefont {Vrhovac}}]{vscepanovic2019PhysicaA}%
  \BibitemOpen
  \bibfield  {author} {\bibinfo {author} {\bibfnamefont {J.}~\bibnamefont {{\v{S}}{\'c}epanovi{\'c}}}, \bibinfo {author} {\bibfnamefont {A.}~\bibnamefont {Kara{\v{c}}}}, \bibinfo {author} {\bibfnamefont {Z.}~\bibnamefont {Jak{\v{s}}i{\'c}}}, \bibinfo {author} {\bibfnamefont {L.}~\bibnamefont {Budinski-Petkovi{\'c}}},\ and\ \bibinfo {author} {\bibfnamefont {S.}~\bibnamefont {Vrhovac}},\ }\bibfield  {title} {\bibinfo {title} {Group chase and escape in the presence of obstacles},\ }\href@noop {} {\bibfield  {journal} {\bibinfo  {journal} {Physica A: Statistical Mechanics and its Applications}\ }\textbf {\bibinfo {volume} {525}},\ \bibinfo {pages} {450} (\bibinfo {year} {2019})}\BibitemShut {NoStop}%
\bibitem [{\citenamefont {Fortunato}(2010)}]{Fortunato2010}%
  \BibitemOpen
  \bibfield  {author} {\bibinfo {author} {\bibfnamefont {S.}~\bibnamefont {Fortunato}},\ }\bibfield  {title} {\bibinfo {title} {Community detection in graphs},\ }\href@noop {} {\bibfield  {journal} {\bibinfo  {journal} {Phys. Rep.-Rev. Sec. Phys. Lett.}\ }\textbf {\bibinfo {volume} {486}},\ \bibinfo {pages} {75} (\bibinfo {year} {2010})}\BibitemShut {NoStop}%
\bibitem [{\citenamefont {Newman}\ and\ \citenamefont {Girvan}(2004)}]{NewmanGirvan2004}%
  \BibitemOpen
  \bibfield  {author} {\bibinfo {author} {\bibfnamefont {M.~E.~J.}\ \bibnamefont {Newman}}\ and\ \bibinfo {author} {\bibfnamefont {M.}~\bibnamefont {Girvan}},\ }\bibfield  {title} {\bibinfo {title} {Finding and evaluating community structure in networks},\ }\href@noop {} {\bibfield  {journal} {\bibinfo  {journal} {Phys. Rev. E.}\ }\textbf {\bibinfo {volume} {69}},\ \bibinfo {pages} {026113} (\bibinfo {year} {2004})}\BibitemShut {NoStop}%
\bibitem [{\citenamefont {Bernstein}\ \emph {et~al.}(2022)\citenamefont {Bernstein}, \citenamefont {Hamblen}, \citenamefont {Junge},\ and\ \citenamefont {Reeves}}]{bernstein2022ECP}%
  \BibitemOpen
  \bibfield  {author} {\bibinfo {author} {\bibfnamefont {E.}~\bibnamefont {Bernstein}}, \bibinfo {author} {\bibfnamefont {C.}~\bibnamefont {Hamblen}}, \bibinfo {author} {\bibfnamefont {M.}~\bibnamefont {Junge}},\ and\ \bibinfo {author} {\bibfnamefont {L.}~\bibnamefont {Reeves}},\ }\bibfield  {title} {\bibinfo {title} {Chase-escape on the configuration model},\ }\href@noop {} {\bibfield  {journal} {\bibinfo  {journal} {Electronic Communications in Probability}\ }\textbf {\bibinfo {volume} {27}},\ \bibinfo {pages} {1} (\bibinfo {year} {2022})}\BibitemShut {NoStop}%
\bibitem [{\citenamefont {Hinsen}\ \emph {et~al.}(2019)\citenamefont {Hinsen}, \citenamefont {Jahnel}, \citenamefont {Cali},\ and\ \citenamefont {Wary}}]{hinsen2019arvix}%
  \BibitemOpen
  \bibfield  {author} {\bibinfo {author} {\bibfnamefont {A.}~\bibnamefont {Hinsen}}, \bibinfo {author} {\bibfnamefont {B.}~\bibnamefont {Jahnel}}, \bibinfo {author} {\bibfnamefont {E.}~\bibnamefont {Cali}},\ and\ \bibinfo {author} {\bibfnamefont {J.-P.}\ \bibnamefont {Wary}},\ }\bibfield  {title} {\bibinfo {title} {Phase transitions for chase-escape models on gilbert graphs},\ }\href@noop {} {\bibfield  {journal} {\bibinfo  {journal} {arXiv preprint arXiv:1911.02622}\ } (\bibinfo {year} {2019})}\BibitemShut {NoStop}%
\bibitem [{\citenamefont {Cali}\ \emph {et~al.}(2024)\citenamefont {Cali}, \citenamefont {Hinsen}, \citenamefont {Jahnel},\ and\ \citenamefont {Wary}}]{cali2024JAP}%
  \BibitemOpen
  \bibfield  {author} {\bibinfo {author} {\bibfnamefont {E.}~\bibnamefont {Cali}}, \bibinfo {author} {\bibfnamefont {A.}~\bibnamefont {Hinsen}}, \bibinfo {author} {\bibfnamefont {B.}~\bibnamefont {Jahnel}},\ and\ \bibinfo {author} {\bibfnamefont {J.-P.}\ \bibnamefont {Wary}},\ }\bibfield  {title} {\bibinfo {title} {Chase--escape in dynamic device-to-device networks},\ }\href@noop {} {\bibfield  {journal} {\bibinfo  {journal} {Journal of Applied Probability}\ }\textbf {\bibinfo {volume} {61}},\ \bibinfo {pages} {311} (\bibinfo {year} {2024})}\BibitemShut {NoStop}%
\bibitem [{\citenamefont {Beckman}\ \emph {et~al.}(2021)\citenamefont {Beckman}, \citenamefont {Cook}, \citenamefont {Eikmeier}, \citenamefont {Hernandez-Torres},\ and\ \citenamefont {Junge}}]{Beckman2021}%
  \BibitemOpen
  \bibfield  {author} {\bibinfo {author} {\bibfnamefont {E.}~\bibnamefont {Beckman}}, \bibinfo {author} {\bibfnamefont {K.}~\bibnamefont {Cook}}, \bibinfo {author} {\bibfnamefont {N.}~\bibnamefont {Eikmeier}}, \bibinfo {author} {\bibfnamefont {S.}~\bibnamefont {Hernandez-Torres}},\ and\ \bibinfo {author} {\bibfnamefont {M.}~\bibnamefont {Junge}},\ }\bibfield  {title} {\bibinfo {title} {Chase-escape with death on trees},\ }\bibfield  {journal} {\bibinfo  {journal} {The Annals of Probability}\ }\textbf {\bibinfo {volume} {49}},\ \href {https://doi.org/10.1214/21-aop1514} {10.1214/21-aop1514} (\bibinfo {year} {2021})\BibitemShut {NoStop}%
\bibitem [{\citenamefont {Wang}\ \emph {et~al.}(2017)\citenamefont {Wang}, \citenamefont {Han},\ and\ \citenamefont {Yang}}]{wang2017PhysicaA}%
  \BibitemOpen
  \bibfield  {author} {\bibinfo {author} {\bibfnamefont {H.}~\bibnamefont {Wang}}, \bibinfo {author} {\bibfnamefont {W.}~\bibnamefont {Han}},\ and\ \bibinfo {author} {\bibfnamefont {J.}~\bibnamefont {Yang}},\ }\bibfield  {title} {\bibinfo {title} {Group chase and escape with sight-limited chasers},\ }\href@noop {} {\bibfield  {journal} {\bibinfo  {journal} {Physica A: Statistical Mechanics and its Applications}\ }\textbf {\bibinfo {volume} {465}},\ \bibinfo {pages} {34} (\bibinfo {year} {2017})}\BibitemShut {NoStop}%
\bibitem [{\citenamefont {Su}\ \emph {et~al.}(2023)\citenamefont {Su}, \citenamefont {Bernardi},\ and\ \citenamefont {Lindner}}]{su2023NJP}%
  \BibitemOpen
  \bibfield  {author} {\bibinfo {author} {\bibfnamefont {M.}~\bibnamefont {Su}}, \bibinfo {author} {\bibfnamefont {D.}~\bibnamefont {Bernardi}},\ and\ \bibinfo {author} {\bibfnamefont {B.}~\bibnamefont {Lindner}},\ }\bibfield  {title} {\bibinfo {title} {Pursuit problem with a stochastic prey that sees its chasers},\ }\href@noop {} {\bibfield  {journal} {\bibinfo  {journal} {New Journal of Physics}\ }\textbf {\bibinfo {volume} {25}},\ \bibinfo {pages} {023033} (\bibinfo {year} {2023})}\BibitemShut {NoStop}%
\bibitem [{\citenamefont {Kaplan}\ and\ \citenamefont {Meier}(1958)}]{kaplan1958TeF}%
  \BibitemOpen
  \bibfield  {author} {\bibinfo {author} {\bibfnamefont {E.~L.}\ \bibnamefont {Kaplan}}\ and\ \bibinfo {author} {\bibfnamefont {P.}~\bibnamefont {Meier}},\ }\bibfield  {title} {\bibinfo {title} {Nonparametric estimation from incomplete observations},\ }\href@noop {} {\bibfield  {journal} {\bibinfo  {journal} {Journal of the American statistical association}\ }\textbf {\bibinfo {volume} {53}},\ \bibinfo {pages} {457} (\bibinfo {year} {1958})}\BibitemShut {NoStop}%
\bibitem [{\citenamefont {Colosimo}\ \emph {et~al.}(2002)\citenamefont {Colosimo}, \citenamefont {Ferreira}, \citenamefont {Oliveira},\ and\ \citenamefont {Sousa}}]{colosimo2002TF}%
  \BibitemOpen
  \bibfield  {author} {\bibinfo {author} {\bibfnamefont {E.}~\bibnamefont {Colosimo}}, \bibinfo {author} {\bibfnamefont {F.~v.}\ \bibnamefont {Ferreira}}, \bibinfo {author} {\bibfnamefont {M.}~\bibnamefont {Oliveira}},\ and\ \bibinfo {author} {\bibfnamefont {C.}~\bibnamefont {Sousa}},\ }\bibfield  {title} {\bibinfo {title} {Empirical comparisons between kaplan-meier and nelson-aalen survival function estimators},\ }\href@noop {} {\bibfield  {journal} {\bibinfo  {journal} {Journal of Statistical Computation and Simulation}\ }\textbf {\bibinfo {volume} {72}},\ \bibinfo {pages} {299} (\bibinfo {year} {2002})}\BibitemShut {NoStop}%
\bibitem [{\citenamefont {Moreira-Soares}\ \emph {et~al.}(2024)\citenamefont {Moreira-Soares}, \citenamefont {Mossmann}, \citenamefont {Travasso},\ and\ \citenamefont {Bordin}}]{moreira2024trajpy}%
  \BibitemOpen
  \bibfield  {author} {\bibinfo {author} {\bibfnamefont {M.}~\bibnamefont {Moreira-Soares}}, \bibinfo {author} {\bibfnamefont {E.}~\bibnamefont {Mossmann}}, \bibinfo {author} {\bibfnamefont {R.~D.}\ \bibnamefont {Travasso}},\ and\ \bibinfo {author} {\bibfnamefont {J.~R.}\ \bibnamefont {Bordin}},\ }\bibfield  {title} {\bibinfo {title} {Trajpy: empowering feature engineering for trajectory analysis across domains},\ }\href@noop {} {\bibfield  {journal} {\bibinfo  {journal} {Bioinformatics Advances}\ }\textbf {\bibinfo {volume} {4}},\ \bibinfo {pages} {vbae026} (\bibinfo {year} {2024})}\BibitemShut {NoStop}%
\bibitem [{\citenamefont {Cormen}\ \emph {et~al.}(2022)\citenamefont {Cormen}, \citenamefont {Leiserson}, \citenamefont {Rivest},\ and\ \citenamefont {Stein}}]{cormen2022MIT}%
  \BibitemOpen
  \bibfield  {author} {\bibinfo {author} {\bibfnamefont {T.~H.}\ \bibnamefont {Cormen}}, \bibinfo {author} {\bibfnamefont {C.~E.}\ \bibnamefont {Leiserson}}, \bibinfo {author} {\bibfnamefont {R.~L.}\ \bibnamefont {Rivest}},\ and\ \bibinfo {author} {\bibfnamefont {C.}~\bibnamefont {Stein}},\ }\href@noop {} {\emph {\bibinfo {title} {Introduction to algorithms}}}\ (\bibinfo  {publisher} {MIT press},\ \bibinfo {year} {2022})\BibitemShut {NoStop}%
\bibitem [{\citenamefont {Hansen}\ \emph {et~al.}(2023)\citenamefont {Hansen}, \citenamefont {Domenici}, \citenamefont {Bartashevich}, \citenamefont {Burns},\ and\ \citenamefont {Krause}}]{Hansen2023}%
  \BibitemOpen
  \bibfield  {author} {\bibinfo {author} {\bibfnamefont {M.~J.}\ \bibnamefont {Hansen}}, \bibinfo {author} {\bibfnamefont {P.}~\bibnamefont {Domenici}}, \bibinfo {author} {\bibfnamefont {P.}~\bibnamefont {Bartashevich}}, \bibinfo {author} {\bibfnamefont {A.}~\bibnamefont {Burns}},\ and\ \bibinfo {author} {\bibfnamefont {J.}~\bibnamefont {Krause}},\ }\bibfield  {title} {\bibinfo {title} {Mechanisms of group‐hunting in vertebrates},\ }\href {https://doi.org/10.1111/brv.12973} {\bibfield  {journal} {\bibinfo  {journal} {Biological Reviews}\ }\textbf {\bibinfo {volume} {98}},\ \bibinfo {pages} {1687–1711} (\bibinfo {year} {2023})}\BibitemShut {NoStop}%
\end{thebibliography}%

\end{document}